\documentclass[twocolumn,twocolappendix]{aastex631}

\usepackage{natbib,aas_macros,amsmath}
\usepackage{multirow,color}
\usepackage{natbib}
\usepackage{multirow}

\def\tcb{\textcolor{blue}}

\newcommand{\oiii}{[O\,{\sc iii}]}
\newcommand{\oii}{[O\,{\sc ii}]}

\newcommand{\nii}{[N\,{\sc ii}]}

\newcommand{\hst}{{\it HST}}
\newcommand{\jwst}{{\it JWST}}

\newcommand{\xion}{$\xi_{\rm ion}$}

\accepted{in ApJL}

\shorttitle{
A NIRSpec First-Look at NIRCam-selected Galaxies at $z\gtrsim8$ in CEERS}
\shortauthors{Fujimoto et al.}

\begin{document}

\title{
CEERS Spectroscopic Confirmation of NIRCam-Selected \boldmath $z\gtrsim8$ Galaxy Candidates \\ 
with \emph{JWST}/NIRSpec: Initial Characterization of their Properties 
}

\correspondingauthor{Seiji Fujimoto}
\email{fujimoto@utexas.edu}
\author[0000-0001-7201-5066]{Seiji Fujimoto}\altaffiliation{Hubble Fellow}
\affiliation{Department of Astronomy, The University of Texas at Austin, Austin, TX, USA}


\author[0000-0002-7959-8783]{Pablo Arrabal Haro}
\affiliation{NSF's National Optical-Infrared Astronomy Research Laboratory, 950 N. Cherry Ave., Tucson, AZ 85719, USA}

\author[0000-0001-5414-5131]{Mark Dickinson}
\affiliation{NSF's National Optical-Infrared Astronomy Research Laboratory, 950 N. Cherry Ave., Tucson, AZ 85719, USA}

\author[0000-0001-8519-1130]{Steven L. Finkelstein}
\affiliation{Department of Astronomy, The University of Texas at Austin, Austin, TX, USA}

\author[0000-0001-9187-3605]{Jeyhan S. Kartaltepe}
\affiliation{Laboratory for Multiwavelength Astrophysics, School of Physics and Astronomy, Rochester Institute of Technology, 84 Lomb Memorial Drive, Rochester, NY 14623, USA}

\author[0000-0003-2366-8858]{Rebecca L. Larson}
\altaffiliation{NSF Graduate Fellow}
\affiliation{Department of Astronomy, The University of Texas at Austin, Austin, TX, USA}

\author[0000-0002-4193-2539]{Denis Burgarella}
\affiliation{Aix Marseille Univ, CNRS, CNES, LAM Marseille, France}


\author[0000-0002-9921-9218]{Micaela B. Bagley}
\affiliation{Department of Astronomy, The University of Texas at Austin, Austin, TX, USA}

\author[0000-0002-2517-6446]{Peter Behroozi}
\affiliation{Department of Astronomy and Steward Observatory, University of Arizona, Tucson, AZ 85721, USA}
\affiliation{Division of Science, National Astronomical Observatory of Japan, 2-21-1 Osawa, Mitaka, Tokyo 181-8588, Japan}

\author[0000-0003-4922-0613]{Katherine Chworowsky}\altaffiliation{NSF Graduate Fellow}
\affiliation{Department of Astronomy, The University of Texas at Austin, Austin, TX, USA}

\author[0000-0002-9921-9218]{Michaela Hirschmann}
\affiliation{Institute for Physics, Laboratory for Galaxy Evolution and Spectral modelling, Ecole Polytechnique Federale de Lausanne, Observatoire de Sauverny, Chemin Pegasi 51, 1290 Versoix,
Switzerland}
\affiliation{INAF, Osservatorio Astronomico di Trieste, Via Tiepolo 11, 34131 Trieste, Italy}

\author[0000-0002-1410-0470]{Jonathan R. Trump}
\affiliation{Department of Physics, 196 Auditorium Road, Unit 3046, University of Connecticut, Storrs, CT 06269, USA}

\author[0000-0003-3903-6935]{Stephen M.~Wilkins} %
\affiliation{Astronomy Centre, University of Sussex, Falmer, Brighton BN1 9QH, UK}
\affiliation{Institute of Space Sciences and Astronomy, University of Malta, Msida MSD 2080, Malta}

\author[0000-0003-3466-035X]{{L. Y. Aaron} {Yung}}
\altaffiliation{NASA Postdoctoral Fellow}
\affiliation{Astrophysics Science Division, NASA Goddard Space Flight Center, 8800 Greenbelt Rd, Greenbelt, MD 20771, USA}


\author[0000-0002-6610-2048]{Anton M. Koekemoer}
\affiliation{Space Telescope Science Institute, 3700 San Martin Drive, Baltimore, MD 21218, USA}

\author[0000-0001-7503-8482]{Casey Papovich}
\affiliation{Department of Physics and Astronomy, Texas A\&M University, College Station, TX, 77843-4242 USA}
\affiliation{George P.\ and Cynthia Woods Mitchell Institute for Fundamental Physics and Astronomy, Texas A\&M University, College Station, TX, 77843-4242 USA}

\author[0000-0003-3382-5941]{Nor Pirzkal}
\affiliation{ESA/AURA Space Telescope Science Institute}


\author[0000-0001-7113-2738]{Henry C. Ferguson}
\affiliation{Space Telescope Science Institute, 3700 San Martin Drive, Baltimore, MD 21218, USA}

\author[0000-0003-3820-2823]{Adriano Fontana}
\affiliation{INAF Osservatorio Astronomico di Roma, Via Frascati 33, 00078 Monte Porzio Catone, Rome, Italy}

\author[0000-0001-9440-8872]{Norman A. Grogin}
\affiliation{Space Telescope Science Institute, 3700 San Martin Drive, Baltimore, MD 21218, USA}

\author[0000-0002-5688-0663]{Andrea Grazian}
\affiliation{INAF--Osservatorio Astronomico di Padova, Vicolo dell'Osservatorio 5, I-35122, Padova, Italy}

\author[0000-0001-8152-3943]{Lisa J. Kewley}
\affiliation{Harvard-Smithsonian Center for Astrophysics, 60 Garden Street, Cambridge, MA 02138, USA}

\author[0000-0002-8360-3880]{Dale D. Kocevski}
\affiliation{Department of Physics and Astronomy, Colby College, Waterville, ME 04901, USA}

\author[0000-0003-3130-5643]{Jennifer M. Lotz}
\affiliation{Gemini Observatory/NSF's National Optical-Infrared Astronomy Research Laboratory, 950 N. Cherry Ave., Tucson, AZ 85719, USA}

\author[0000-0001-8940-6768]{Laura Pentericci}
\affiliation{INAF Osservatorio Astronomico di Roma, Via Frascati 33, 00078 Monte Porzio Catone, Rome, Italy}

\author[0000-0002-5269-6527]{Swara Ravindranath}
\affiliation{Space Telescope Science Institute, 3700 San Martin Drive, Baltimore, MD 21218, USA}

\author[0000-0002-6748-6821]{Rachel S.~Somerville}
\affiliation{Center for Computational Astrophysics, Flatiron Institute, 162 5th Avenue, New York, NY 10010, USA}

\author[0000-0003-3903-6935]{Stephen M.~Wilkins} %
\affiliation{Astronomy Centre, University of Sussex, Falmer, Brighton BN1 9QH, UK}
\affiliation{Institute of Space Sciences and Astronomy, University of Malta, Msida MSD 2080, Malta}


\author[0000-0001-5758-1000]{Ricardo O. Amor\'{i}n}
\affiliation{Instituto de Investigaci\'{o}n Multidisciplinar en Ciencia y Tecnolog\'{i}a, Universidad de La Serena, Raul Bitr\'{a}n 1305, La Serena 2204000, Chile}
\affiliation{Departamento de Astronom\'{i}a, Universidad de La Serena, Av. Juan Cisternas 1200 Norte, La Serena 1720236, Chile}

\author[0000-0001-8534-7502]{Bren E. Backhaus}
\affiliation{Department of Physics, 196 Auditorium Road, Unit 3046, University of Connecticut, Storrs, CT 06269, USA}

\author[0000-0003-2536-1614]{Antonello Calabr{\`o}}
\affiliation{INAF Osservatorio Astronomico di Roma, Via Frascati 33, 00078 Monte Porzio Catone, Rome, Italy}

\author[0000-0002-0930-6466]{Caitlin M.\ Casey}
\affiliation{Department of Astronomy, The University of Texas at Austin, Austin, TX, USA}

\author[0000-0003-1371-6019]{M. C. Cooper}
\affiliation{Department of Physics \& Astronomy, University of California, Irvine, 4129 Reines Hall, Irvine, CA 92697, USA}

\author[0000-0003-0531-5450]{Vital Fern\'andez}
\affiliation{Departamento de Astronom\'{i}a, Universidad de La Serena, Av. Juan Cisternas 1200 Norte, La Serena 1720236, Chile}

\author[0000-0002-3560-8599]{Maximilien Franco}
\affiliation{Department of Astronomy, The University of Texas at Austin, Austin, TX, USA}

\author[0000-0002-7831-8751]{Mauro Giavalisco}
\affiliation{University of Massachusetts Amherst, 710 North Pleasant Street, Amherst, MA 01003-9305, USA}

\author[0000-0001-6145-5090]{Nimish P. Hathi}
\affiliation{Space Telescope Science Institute, 3700 San Martin Drive, Baltimore, MD 21218, USA}

\author[0000-0003-0129-2079]{Santosh Harish}
\affiliation{Laboratory for Multiwavelength Astrophysics, School of Physics and Astronomy, Rochester Institute of Technology, 84 Lomb Memorial Drive, Rochester, NY 14623, USA}

\author[0000-0001-6251-4988]{Taylor A. Hutchison}
\altaffiliation{NASA Postdoctoral Fellow}
\affiliation{Astrophysics Science Division, NASA Goddard Space Flight Center, 8800 Greenbelt Rd, Greenbelt, MD 20771, USA}

\author[0000-0001-9298-3523]{Kartheik G. Iyer}
\affiliation{Dunlap Institute for Astronomy \& Astrophysics, University of Toronto, Toronto, ON M5S 3H4, Canada}

\author[0000-0003-1187-4240]{Intae Jung}
\affiliation{Space Telescope Science Institute, 3700 San Martin Drive, Baltimore, MD 21218, USA}

\author[0000-0003-1581-7825]{Ray A. Lucas}
\affiliation{Space Telescope Science Institute, 3700 San Martin Drive, Baltimore, MD 21218, USA}

\author[0000-0002-0786-7307]{Jorge A. Zavala}
\affiliation{Division of Science, National Astronomical Observatory of Japan, 2-21-1 Osawa, Mitaka, Tokyo 181-8588, Japan}

\begin{abstract}
We present \jwst\ NIRSpec spectroscopy for 11 galaxy candidates 
with photometric redshifts of $z\simeq9-13$ and 
$M_{\rm UV} \in[-21,-18]$ newly identified in NIRCam images in the Cosmic Evolution Early Release Science (CEERS) Survey. 
We confirm emission line redshifts for 7 galaxies at $z=7.762$--8.998 using spectra at $\sim1$--5$\mu$m either with the NIRSpec prism or its three medium resolution ($R \sim$ 1000) gratings.
For $z\simeq9$ photometric candidates, we achieve a high confirmation rate of $\simeq$ 90\%, which validates the classical dropout selection from NIRCam photometry. 
No robust emission lines are identified in three galaxy candidates at $z>10$, where the strong [O{\sc iii}] and H$\beta$ lines would be redshifted beyond the wavelength range observed by NIRSpec, and the Lyman-$\alpha$ continuum break is not detected with the sensitivity of the current data.
Compared with \hst-selected bright galaxies ($M_{\rm UV}\simeq-22$) that are similarly spectroscopically confirmed at $z\simeq8-9$,
these NIRCam-selected galaxies are characterized 
by lower star formation rates (SFR $\simeq4\,M_{\odot}$~yr$^{-1}$) and lower stellar masses ($\simeq10^{8}\, M_{\odot}$), but with higher specific SFR ($\simeq$ 40~Gyr$^{-1}$), higher \oiii+H$\beta$ equivalent widths ($\simeq1100 {\rm \AA}$), and elevated production efficiency of ionizing photons ($\log(\xi_{\rm ion}/{\rm Hz \,erg}^{-1})\simeq25.8$) induced by young stellar populations ($<10$ Myrs) accounting for $\simeq20\%$ of the galaxy mass, 
highlighting the key contribution of faint galaxies to cosmic reionization. 
Taking advantage of the homogeneous selection and sensitivity, 
we also investigate metallicity and ISM conditions with empirical calibrations using the \oiii$_{5008}$/H$\beta$ ratio. 
We find that galaxies at $z \simeq8-9$ have higher SFRs and lower metallicities than galaxies at similar stellar masses at $z \simeq2-6$,
which is generally consistent with the current galaxy formation and evolution models. 
\end{abstract}

\section{Introduction} \label{sec:intro}

Studying early galaxies is key to understanding fundamental cosmological questions such as the development of large-scale structure, dark matter, and the processes that govern cosmic reionization and early galaxy formation and evolution. 
In the last decades, the search for galaxies seen within the epoch of Reionization (EoR) has been successful at $6\lesssim z\lesssim11$. With thousands of galaxies discovered, 
deep \textit{Hubble Space Telescope (HST)} surveys have provided valuable demographic data for these galaxies, including an initial characterization of the stellar component, in terms of un-obscured star formation rates and sizes \citep[e.g.,][]{ellis2013, bouwens2015, finkelstein2015, oesch2016, bhatawdekar2019}.

The start of \textit{James Webb Space Telescope} (\textit{JWST}) operations \citep{rigby2022} has led to significant progress in the discovery and investigation of galaxies at very early cosmic epochs. 
From the Early Release Observations (ERO; \citealt{pontoppidan2022}) and the Early Release Science programs (ERS; e.g., \citealt{treu2022, finkelstein2023}),  
multiple NIRCam imaging surveys have been carried out towards both lensing clusters and blank fields, 
where dozens of high-redshift galaxy candidates have been 
identified at $z\simeq$ 9--17 \citep[e.g.,][]{adams2022, atek2022, bouwens2022c, bradley2022, castellano2022, donnan2023, finkelstein2022b, finkelstein2023, harikane2023, labbe2022, morishita2022, naidu2022, yan2022}. 
Their abundance at the bright-end ($M_{\rm UV} \lesssim -20$) exceeds nearly all theoretical predictions so far \citep[e.g.,][]{behroozi2015, dayal2017,   yung2019a, yung2020b, behroozi2019, behroozi2020, dave2019, wilkins2022a, wilkins2022b, kannan2022,  mason2022}, 
suggesting several possibilities, including that star formation in early systems is dominated by a top-heavy initial mass function (IMF), complete lack of dust attenuation, and/or changing star-formation physics \citep[e.g.,][]{harikane2023, finkelstein2023, ferrara2022, boylan2022, lovell2022, menci2022}. 

Deep ALMA follow-up observations have been immediately performed through Director’s Discretionary Time (DDT) for several remarkably bright ($M_{\rm UV}\in[-22,-21]$) \jwst\ galaxy candidates at $z\sim11$--17 \citep[e.g.,][]{castellano2022, naidu2022, harikane2023}, showing no robust dust continuum detection from any of these candidates \citep{bakx2022, yoon2022, fujimoto2022b}. 
These ALMA results disfavor the possibility of lower-$z$ dusty interlopers \citep{fujimoto2022b} and show possible  \oiii\ line detection \citep{bakx2022, yoon2022, fujimoto2022b}, although robust spectroscopic confirmation of these sources is inevitably required. 

In this paper, we present \jwst/NIRSpec results of follow-up observations of $z\gtrsim$ 8.5 galaxy candidates identified in the first epoch of the Cosmic Evolution Early Release Science (CEERS) Survey. 
This is the first homogeneous, luminosity-selected follow-up spectroscopy for \jwst\ high-redshift galaxy candidates at $z\gtrsim8.5$ in the UV luminosity range of $M_{\rm UV} \in[-21,-18]$. 
This sets the benchmark for the spectroscopic confirmation of large samples of galaxies, further allowing detailed investigations into the UV luminosity function shape at $z\gtrsim8.5$ and the characterization of the high-redshift Lyman-break galaxy (LBG) population newly identified with \jwst.  This work is complementary to the recent successful spectroscopic confirmation of the higher redshift, but much (1--2 mag) fainter galaxies at $z=9.51$--13.20 in lensing cluster fields \citep{roberts-borsani2022b, williams2022} or very deep observations of the Hubble Ultra Deep Field \citep{robertson2022, curtis-lake2022}. 

The structure of this paper is as follows. 
In Section 2, we describe the observations and the data reduction of CEERS \jwst\ NIRCam and NIRSpec observations. 
Section 3 outlines the methods of redshift determination. 
In Section 4, we present and discuss the results of the validity of the redshift determination and the physical properties for the spectroscopically confirmed sources. 
A summary of this study is presented in Section 5. 
Throughout this paper, we assume a flat universe with 
$\Omega_{\rm m} =$ 0.3, 
$\Omega_\Lambda =$ 0.7, 
and $H_0 =$ 70 km s$^{-1}$ Mpc$^{-1}$ 
and the \cite{chabrier2003} initial mass function (IMF). 
We adopt vacuum rest-frame wavelengths for the emission lines. 
When we show lensed galaxies taken from the literature for comparison, we show the intrinsic properties after magnification corrections. 

\begin{figure*}[t]
\begin{center}
\includegraphics[trim=0cm 0cm 0cm 0cm, clip, angle=0,width=1.\textwidth]{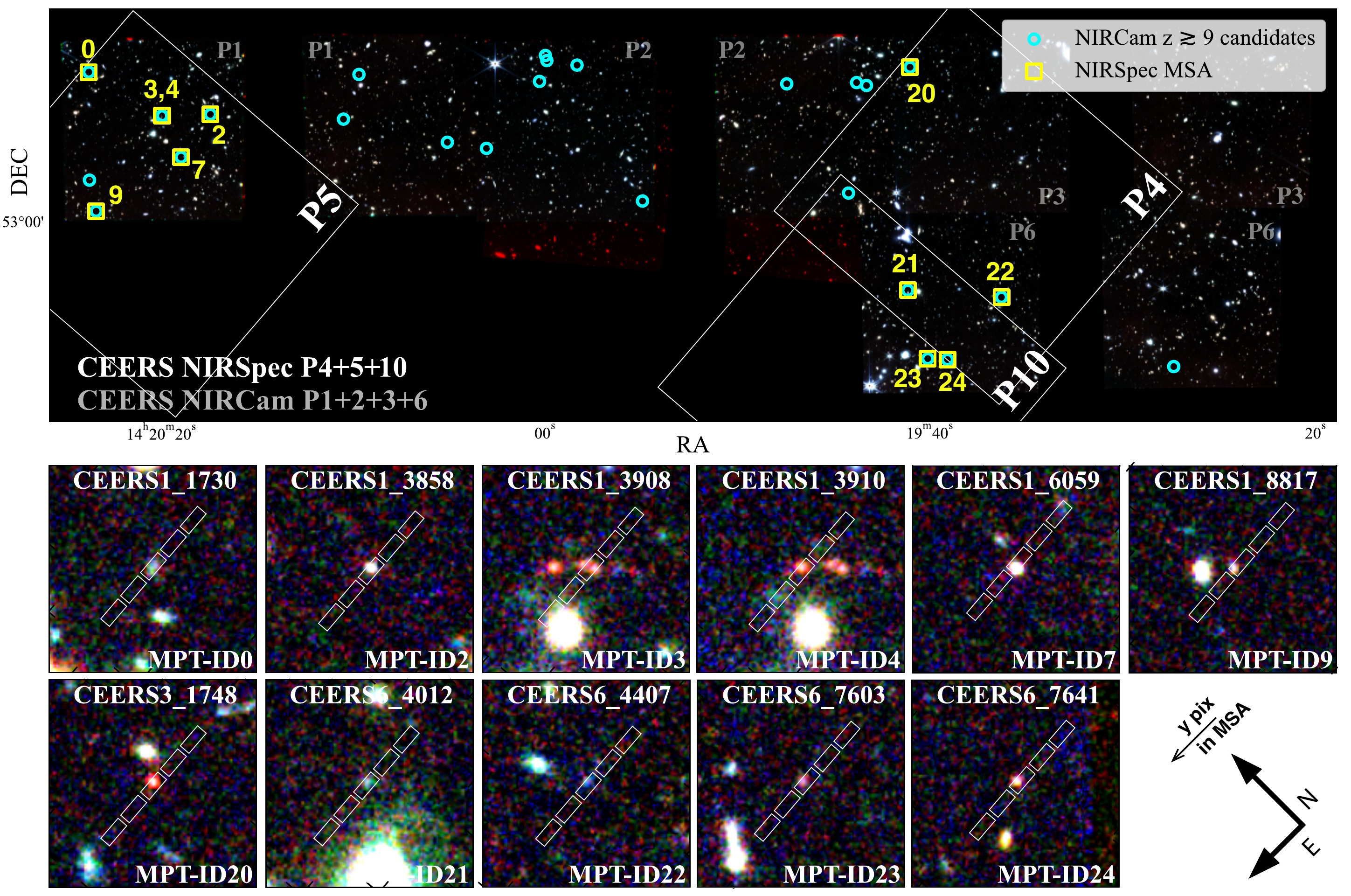}
 \caption{
\textit{\textbf{Top:}} A NIRCam RGB color (R: F444W, G: F356W, B: F277W) mosaic combining the NIRCam pointing IDs of 1, 2, 3, and 6 taken in CEERS epoch 1 (June 2022). 
The white squares show the NIRSpec MSA footprints of the NIRSpec pointing IDs of 4, 5, and 10 taken in CEERS epoch 2 (December 2022). 
The cyan circles denote the NIRCam-selected high-redshift galaxy candidates at $z\gtrsim8.5$ presented in \cite{finkelstein2023}. 
The yellow squares represent the NIRSpec MSA targets, and their MPT IDs are labeled.  
Two candidates in the MSA footprints did not receive slits due to the constraints of the MSA configuration. 
\textit{\textbf{Bottom:}} 
RGB $3\farcs6\times3\farcs6$ image cutouts around the MSA targets (R: F356W, G: F277W, B: F150W). 
The rectangles show the $0\farcs2\times0\farcs46$ shutter configuration.  
We use the standard 3-shutter MSA slitlets and perform a 3-point nodding. 
Thus, five shutter positions are presented, including the nod positions, which is reflected in the 2D spectra in Figure~\ref{fig:spectra}. 
\label{fig:footprint}}
\end{center}
\end{figure*}

\setlength{\tabcolsep}{6pt}
\begin{table*}[t]
\begin{center}
\caption{NIRSpec follow-up high-$z$ targets identified with NIRCam in CEERS epoch 1}
\label{tab:target}
\vspace{-0.4cm}
\begin{tabular}{cccccccccc}
\hline  \hline
Source ID$^{\dagger}$ & MPT-ID & R.A.   & Dec. & $m_{\rm F277W}$ & $z_{\rm phot}$ & $z_{\rm spec}$ & S/N & Mode & Other Ref. \\ 
         &   & (deg) & (deg) &      (mag)     &   &                          &                &   &   \\  
    (1)     & (2)    & (3)    & (4)            &       (5)                   & (6)            & (7) & (8) & (9) & (10) \\  \hline  
   \multicolumn{10}{c}{Confirmed} \\ \hline
CEERS1\_3858 & 2 & 214.994402 &  52.989379 &  $27.2$ & $8.95^{+0.15}_{-0.18}$  & $8.807\pm0.003$ & 7.5 & P  & B22, D23, W23 \\ 
CEERS1\_3908 & 3 & 215.005189 &  52.996580 &  $27.3$ & $9.04^{+1.29}_{-0.06}$  & $8.005\pm0.001$$^{\dagger\dagger}$ & 17.5 & P & D23 \\ 
CEERS1\_3910 &  4 & 215.005365 &  52.996697 &  $28.0$ & $9.55^{+1.05}_{-0.39}$  & $7.9932\pm0.0006$ & 9.6 &M & B22, D23, L23 \\ 
CEERS1\_6059 & 7 & 215.011706 &  52.988303 &  $27.0$ & $9.01^{+0.06}_{-0.06}$ & $8.876\pm0.002$ & 8.9 &P,M & D23, W23 \\ 
CEERS3\_1748 & 20 & 214.830685 &  52.887771 &  $28.5$ & $8.77^{+0.45}_{-1.08}$ & $7.769\pm0.003$ & 7.7 &P,M & D23, L23 \\ 
CEERS6\_7603 & 23 &214.901252 &  52.846997 &  $28.9$ & $11.32^{+0.30}_{-1.74}$ & $8.8805\pm0.0005$$^{\dagger\dagger}$ & 9.8 & M & D23\\ 
CEERS6\_7641 &  24 & 214.897232 &  52.843854 &  $28.1$ & $8.95^{+1.95}_{-0.15}$ & $8.9980\pm0.0005$$^{\dagger\dagger}$ & 11.1 &M & B22, D23, W23\\ \hline 
\multicolumn{10}{c}{Not confirmed} \\ \hline
CEERS1\_1730 & 0 & 215.010022 &  53.013641 &  $27.7$ & $13.36^{+0.84}_{-1.08}$ & -- & -- & P & \\ 
CEERS6\_4012 & 21 & 214.888127 &  52.858987 &  $27.6$ & $8.89^{+0.36}_{-0.36}$  & -- & -- &M & D23 \\ 
CEERS6\_4407  & 22 & 214.869661 &  52.843646 &  $29.0$ & $10.63^{+0.81}_{-0.57}$ & -- & -- &M & \\ \hline
\multicolumn{10}{c}{Tentative} \\ \hline 
CEERS1\_8817 & 9 & 215.043999 &  52.994302 &  $28.1$ & $10.60^{+0.42}_{-0.36}$ & 9.388$?$ or 9.696$?$ & 5.7 &P & D23 \\ 
\hline \hline
\end{tabular}
\end{center}
\vspace{-0.4cm}
\tablecomments{
(1) Source ID used in \cite{finkelstein2023}. 
(2) Source ID in the NIRSpec MSA configurations. 
(3) Right ascension (J2000). 
(4) Declination (J2000).  
(5) Observed AB total magnitude in the F277W filter. 
(6) Photometric redshift. 
(7) Spectroscopic redshift measured from NIRSpec spectroscopy. 
(8) S/N from the Gaussian+continuum fit for the \oiii5008, 4960 lines. For MPT-ID9, we show the single-line S/N integrated over the 3 pixels around the peak. 
(9) NIRSpec observation mode (P: prism, M: Medium resolution gratings).
(10) References that also report the source identification as high-redshift (\tcb{$z\gtrsim8.5$}) galaxy candidates (B22: \citealt{bouwens2022d}, D23: \citealt{donnan2023}, L23: \citealt{labbe2022}, and W23: \citealt{whitler2023}).
(3)--(6) are taken from \cite{finkelstein2023}. \\
$\dagger$ From top to bottom, these sources are referred to as CEERS\_4702, 4774, 4777, 7078, 23084, 61381, 61419, 2067, 56878, 57400, and 10332 in \cite{arrabal-halo2023b}. \\
$\dagger\dagger$ These three galaxies have also been reported in \cite{tang2023}. 
}
\end{table*}

\section{Observations \& Data Processing} 
\label{sec:data}

The \textit{JWST}/NIRCam \citep{rieke2003,rieke2005,beichman2012} and \textit{JWST}/NIRSpec \citep{Jakobsen2022} data employed in this work were taken as part of the Cosmic Evolution Early Release Science Survey (CEERS; ERS 1345, PI: S. Finkelstein) in the CANDELS \citep{grogin2011, koekemoer2011} Extended Groth Strip (EGS) field. 
The complete details of the CEERS program will be presented in Finkelstein et al., (in prep.).

\subsection{\textit{JWST}/NIRCam data and Target Selection}
\label{sec:NIRCam_data}

The galaxies discussed in this work are part of the photometrically-selected $z\gtrsim8.5$ candidate galaxy sample assembled by \cite{finkelstein2023}.  These galaxy candidates were selected from the CEERS epoch 1 (June 2022) NIRCam imaging data, which are described in full in \citet{Bagley2022}. The \cite{finkelstein2023} photometry catalog includes measurements over the full NIRCam wavelength range in the F115W, F150W, F200W, F277W, F356W, F410M and F444W filters, which have exposures of $\sim3000$ seconds per filter ($\sim$ 6000 s for F115W); as well as in the existing {\it HST}/CANDELS ACS and WFC3 F606W, F814W, F105W, F125W, F140W and F160W bands.  

Robust candidate $z\gtrsim8.5$ galaxies were selected based on a combination of detection significance and photometric redshift distribution criteria, designed to select well-measured astrophysical sources with photometric redshifts highly likely to be at $z\gtrsim8.5$.  This full sample consists of 26 candidate galaxies, with the full details available in \cite{finkelstein2023}. The NIRSpec multi-object spectroscopy (MOS) configurations were designed to maximize the number of these candidates observed, resulting in a total of 11 targets at $z\simeq9$--13.
In Figure~\ref{fig:footprint}, we present the source positions of the full 26 candidates and the 11 targets included in the NIRSpec observations. The target properties are summarized in Table~\ref{tab:target}.

\subsection{\textit{JWST}/NIRSpec data}
\label{sec:NIRSpec_data}

The $z\gtrsim8.5$ candidates presented in this work are included in the NIRSpec MOS configurations taken with the Micro Shutter Array \citep[MSA;][]{Ferruit2022} during the CEERS epoch 2 observations (December 2022). These NIRSpec observations are split into 6 different MSA pointings, each of them observed with the G140M/F100LP, G235M/F170LP and G395M/F290LP medium resolution ($R\approx1000$; here denoted by ``M'') gratings plus the prism ($R\approx30$--300), fully covering the $\sim1$--5 $\mu$m wavelength range. The MSA was configured to use 3-shutter slitlets, enabling a 3-point nodding pattern, shifting the pointing by a shutter length in each direction for background subtraction. The total exposure time per disperser is 3107~s distributed as three integrations (one per nod) of 14 groups each in the NRSIRS2 readout mode. 
Two prism observations (NIRSpec pointings 9 and 10) were affected by an electrical short\footnote{Webb Observing Problem Report (WOPR) ID: 88650}, and we do not include those data in our analyses in this paper. Those two prism observations were rescheduled in CEERS epoch 3 (February 2023), for which the data processing, analyses, and results are presented in \cite{arrabal-halo2023b}.

The M grating and prism observations have different MSA configurations and so not all targets are observed in both modes. 
We present spectra of 11 candidates at $z\gtrsim8.5$ in this study: 7 were observed with the prism, 5 were observed with the M gratings, and two were observed with both modes (see Table~\ref{tab:target}).

\subsection{NIRSpec data reduction}
\label{sec:NIRSpec_reduction}

The details of the CEERS NIRSpec data processing will be presented in Arrabal Haro et al., (in prep.). We summarize the main steps of the reduction here.
The NIRSpec data is processed with the STScI Calibration Pipeline\footnote{\url{https://jwst-pipeline.readthedocs.io/en/latest/index.html}} version 1.8.5 and the Calibration Reference Data System (CRDS) mapping 1027. We make use of the \texttt{calwebb\_detector1} pipeline module to subtract the bias and the dark current, correct the 1/$f$ noise and generate count-rate maps (CRMs) from the uncalibrated images. At this stage, the parameters of the \texttt{jump} step are modified for an improved correction of the ``snowball'' 
events\footnote{\url{https://jwst-docs.stsci.edu/data-artifacts-and-features/snowballs-and-shower-artifacts}} associated with high-energy cosmic rays.

The resulting CRMs are then processed with the \texttt{calwebb\_spec2} pipeline module which creates two-dimensional (2D) cutouts of the slitlets, performs the background subtraction making use of the 3-nod pattern, corrects the flat-fields, implements the wavelength and photometric calibrations and resamples the 2D spectra to correct the distortion of the spectral trace. The \texttt{pathloss} step accounting for the slit loss correction is turned off at this stage of the reduction process. Instead, when required for the analysis, we introduce slit loss corrections based on the morphology of the sources in the NIRCam bands and the location of the slitlet hosting them.

The images of the three nods are combined at the \texttt{calwebb\_spec3} pipeline stage, making use of customized apertures for the extraction of the one-dimensional (1D) spectrum. The custom extraction apertures are visually defined for targets presenting high signal-to-noise ratio (S/N) at the continuum or emission lines, which are easily recognizable in the 2D spectra. For those cases where the targets are too faint for a robust visual identification, a 4-pixel extraction aperture is defined around a spatial location estimated from the relative position of the target within its shutter, derived from the MSA configuration.

Finally, the 2D and 1D spectra are simultaneously inspected with the Mosviz visualization tool\footnote{\url{https://jdaviz.readthedocs.io/en/latest/mosviz/index.html}} \citep{JDADF2023} to mask possible remaining hot pixels and other artifacts in the images, as well as the detector gap (when present). Once inspected and masked, the three M gratings are combined into a single spectrum, resampling to a common wavelength array at the overlapping wavelengths and adopting the mean flux at each pixel, weighted by the flux errors. 
No resampling of the wavelength grid is performed in the prism spectra.

We test the accuracy of the error spectrum by comparing the normalized median absolute deviation of the science spectrum (masking out emission lines) to the median of the error spectrum.  We find that the actual data show fluctuations $\sim$1.5--2$\times$ larger than the typical error value. Thus, we scale the error spectrum up by this scale factor on an object-by-object basis. 
Our noise correction factor from the above method is consistent with the estimates presented in \cite{arrabal-halo2023b}.
Although the effect of noise correlation might still increase the correction factor, any potential impact would likely be modest ($\sim$30\%; \citealt{arrabal-halo2023b}).

\begin{figure*}[t]
\begin{center}
\includegraphics[trim=0.4cm 0cm 0.4cm 0cm, clip, angle=0,width=1.\textwidth]{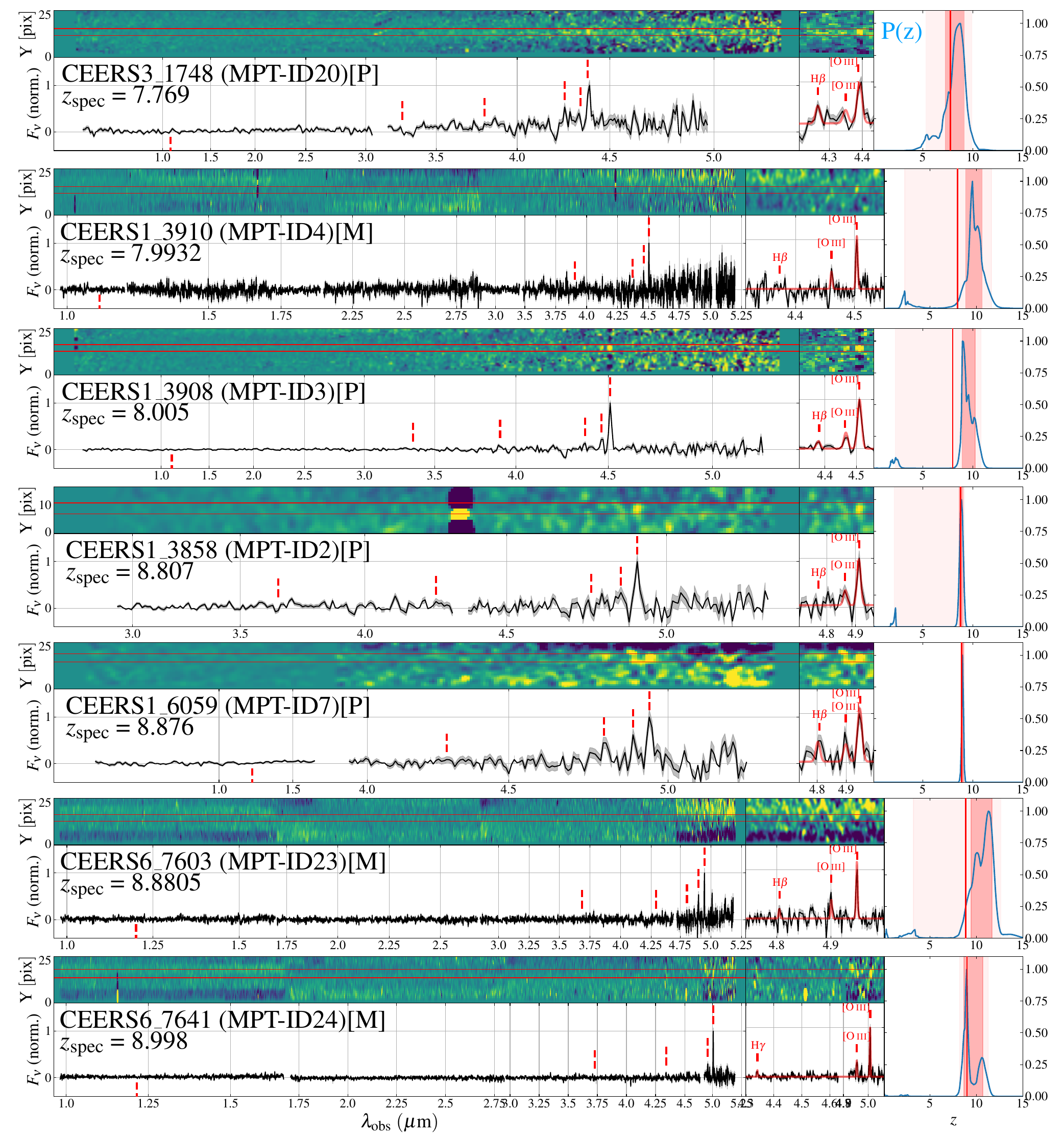}
 \caption{
NIRSpec 2D (top) and 1D (bottom) spectra in the observed wavelength frame for the sources that are spectroscopically confirmed. MPT-ID~20, 3, 2, and 7 show the prism spectrum, while ID~4, 23, and 24 show the combined spectra of the G140M, G235M and G395M gratings.  
The red horizontal lines on the 2D spectrum represent the aperture position used for extracting the 1D spectrum. 
The red vertical dashed lines denote the expected wavelengths of Ly$\alpha$, \oii3727, H$\gamma$, H$\beta$, \oiii4960, and \oiii5008 from left to right based on our $z_{\rm spec}$ measurement. 
The 2D spectrum is smoothed with a one-pixel (=sigma) Gaussian kernel. 
The middle panel shows the zoom-in spectrum around the brightest line feature. 
The best-fit Gaussian + continuum fit is overlaid on the 1D spectrum. 
The right panel presents the photometric redshift probability distribution $P(z)$ as the blue curve, while the red vertical line represents the $z_{\rm spec}$ value.  
The dark and light red shaded regions indicate 68\% and 95\% confidence intervals of $P(z)$, respectively, from the photometric redshift analysis \citep{finkelstein2023}. For sources observed in both the P (prism) and M (medium grating) modes, we show here the spectrum with the better S/N in the brightest emission line. 
\label{fig:spectra}}
\end{center}
\end{figure*}


\section{Analysis}
\label{sec:analysis}

\subsection{Line Identification}
\label{sec:line}

We systematically analyze the reduced 1D and 2D spectra of the 11 spectroscopically-observed targets to identify emission line and continuum features. 
From 7 targets (MPT-ID2, 3, 4, 7, 20, 23, 24), we identify at least two emission line features whose wavelength separations and flux ratios match those of the \oiii5008, 4960 emission lines.  
These line features have S/N $\gtrsim7$ and S/N $\gtrsim3$ from the brighter and the fainter doublet line, respectively, based on the integrated pixel S/N over 3-pixels around the peak. 
In Figure~\ref{fig:spectra}, we show the 1D and 2D spectra of these seven targets.  

From one additional target (MPT-ID9), a single line feature is observed with S/N $>5$ at 5.20~$\mu$m, suggesting a potential identification of \oiii5008 at $z=9.388$ or H$\beta$ at $z=9.696$.  
However, no positive signals at the expected wavelengths for the neighboring lines are confirmed in those potential redshift solutions with H$\beta$, \oiii5008, or \oiii4960. 
This suggests that the neighboring lines are buried in the noise fluctuations or that the 5.20~$\mu$m line feature is just noise. 
We thus focus on the first seven sources with at least two significant line features in the following analyses. From the other three targets (MPT-ID0, 21, 22), we identify no secure line or Lyman-$\alpha$ break features. Interestingly, three out of these four potential and non-detected sources are estimated to have $z_{\rm phot} > 10$ \citep{finkelstein2023}, suggesting that these sources might be truly located at $z>10$, where the typical strong rest-frame optical emission lines (H$\beta$, \oiii5008, 4960) 
would lie beyond the red limit of NIRSpec wavelength coverage, making the redshift estimate challenging.  
The 1D and 2D spectra for these four targets are shown in the Appendix. 

\subsection{Redshift Measurement}
\label{sec:z}

For the seven sources with at least two possible line features, first, we perform template fitting to the NIRSpec 1D spectra to obtain initial redshift estimates. A linear combination of three different {\sc cigale} \citep{burgarella2005, noll2009, boquien2019} models is fitted to the real spectra at a redshift interval, with steps matching the spectral resolution of the grating (or the highest spectral resolution at the red wavelengths in the case of the prism). The templates employed are selected to represent a wide range of galaxy spectra, from emission lines galaxies to quiescent and old massive galaxies. For all the sources spectroscopically confirmed, the best solution from $\chi^{2}$ minimization over the explored $z=$0--13 range is consistent with an \oiii5008 detection at long wavelengths.

Using the best-matching template redshifts as the starting estimation, we perform a Gaussian + continuum fitting with a Markov Chain Monte Carlo (MCMC) approach around the peak of the brightest emission line to refine the redshift measurements. Because these prior redshift estimates suggest that the brightest emission line corresponds to \oiii5008 in all seven sources, we adopt the vacuum rest-frame wavelengths of 5008.22~${\rm \AA}$ and 4960.28~${\rm \AA}$ for the fit to the brightest and neighboring lines simultaneously by assuming that their line widths and the underlying continuum are the same and the line ratio of 3:1.  
We determine spectroscopic redshifts in the range $z_{\rm spec}=7.7621$--8.9979. 
We summarize our final $z_{\rm spec}$ measurements and 1$\sigma$ uncertainties in Table~~\ref{tab:target}.  
In the right panel of Figure \ref{fig:spectra}, the red line and blue curve show our $z_{\rm spec}$ value compared to the photometric redshift probability distribution $P(z)$ presented in \cite{finkelstein2023} based on the photometric SED analysis, respectively. We confirm that the $z_{\rm spec}$ values of all these seven sources are consistent with the $P(z)$ distribution within 2$\sigma$. We further discuss the difference between $z_{\rm spec}$ and $z_{\rm phot}$ in Section~\ref{sec:zspec}.  
We detect the Lyman-$\alpha$ continuum break only from the prism spectrum of MPT-ID7, where we confirm the wavelength of the continuum break is consistent with the redshift determined by the emission lines. 
We show the Lyman-$\alpha$ break observed in MPT-ID7 in the Appendix.

Note that MPT-ID3 and ID4 are separated by less than 3~kpc in physical scale (see Figure~\ref{fig:footprint}), and our $z_{\rm spec}$ measurements are $z=8.005 \pm 0.001$ for ID3 and $z=7.9932 \pm 0.0006$ for ID4. 
Although the line feature is less evident in ID4, 
these independent results suggest that ID3 and ID4 are a close pair 
and our $z_{\rm spec}$ estimate for ID4 is secure. 
We also note that the \oiii5008 line feature in the 2D spectrum of MPT-ID2 is less clear than for other sources. However, we confirm that the S/N increases to 7.5 in the simultaneous fit to both \oiii\ lines by fixing the wavelength separation and the line ratio. 
The $z_{\rm spec}$ value also shows an excellent agreement with $z_{\rm phot}$. 
We thus include MPT-ID2 in our $z_{\rm spec}$ sample in this paper. 

We also run the template fitting for the remaining four sources in order not to miss the chance that we identify multiple marginal line detections whose wavelength separation matches with a specific redshift solution. However, we do not find any 
convincing redshift solutions for these sources.

\setlength{\tabcolsep}{3pt}
\begin{table*}[t!]
\begin{center}
\caption{Physical Properties of the NIRCam CEERS galaxies confirmed at $z\simeq8-9$}
\label{tab:prop}
\vspace{-0.4cm}
\begin{tabular}{cccccccccccc}
\hline  \hline
MPT-ID & $z_{\rm spec}$ & $M_{\rm UV}$ & $E(B-V)$ &   SFR   & $M_{\rm star}$ & $f_{\rm burst}$ & EW([OIII]5008) & EW(H$\beta$) & R3  & O32  & Mode \\
    &                    &  (mag)      &   (mag) & ($M_{\odot}$~yr$^{-1}$) &   ($10^{8}M_{\odot}$) & & (${\rm \AA}$) & (${\rm \AA}$) &     &      &           \\ 
(1)            &            (2)        &  (3)            &        (4)      &  (5) &  (6)     &   (7) & (8) &  (9)   &   (10)   &    (11)  & (12)      \\  \hline 
2 & 8.807 &  $-20.44$ & $0.03^{+0.03}_{-0.02}$ &$3.2^{+1.1}_{-0.5}$ & $1.2^{+0.8}_{-0.6}$ &$0.12^{+0.18}_{-0.08}$ &$ 372\pm112$ &$ < 150$ &$ >2.5$ &$ >8.2$ &P \\
3 & 8.005 &  $-20.47$ & $0.15^{+0.04}_{-0.07}$ &$9.8^{+2.9}_{-3.2}$ & $2.0^{+1.7}_{-0.8}$ &$0.25^{+0.21}_{-0.16}$ &$ 1022\pm129$ &$ 163\pm62$ &$ 6.3\pm2.3$ &$ >8.1$ &P \\
4 & 7.9932 &  $-19.44$ & $0.33^{+0.12}_{-0.09}$ &$16.3^{+21.0}_{-8.1}$ & $8.7^{+6.3}_{-4.4}$ &$0.10^{+0.18}_{-0.07}$ &$ 430\pm69$ &$ < 166$ &$ >2.6$ & \nodata$^{\dagger}$ &M \\
7 & 8.876 &  $-20.75$ & $0.03^{+0.03}_{-0.02}$ &$3.9^{+1.1}_{-0.6}$ & $1.2^{+0.9}_{-0.6}$ &$0.13^{+0.25}_{-0.09}$ &$ 895\pm436$ &$ 350\pm194$ &$ 2.6\pm0.8$ &$ >5.5$ &M \\
20 & 7.769 &  $-18.55$ & $0.61^{+0.02}_{-0.18}$ &$64.3^{+18.6}_{-50.8}$ & $30.6^{+19.7}_{-13.0}$ &$0.08^{+0.13}_{-0.06}$ &$ 109\pm19$ &$ 50\pm16$ &$ 2.2\pm0.8$ &$ >6.1$ &P \\
23 & 8.8805 &  $-18.38$ & $0.06^{+0.1}_{-0.04}$ &$0.8^{+0.7}_{-0.3}$ & $0.2^{+0.2}_{-0.1}$ &$0.21^{+0.23}_{-0.14}$ &$ 1195\pm200$ &$ 208\pm121$ &$ 5.8\pm3.3$ &$ >6.5$ &M \\
24 & 8.998 &  $-19.08$ & $0.09^{+0.08}_{-0.06}$ &$2.7^{+1.5}_{-1.0}$ & $0.6^{+0.6}_{-0.3}$ &$0.23^{+0.22}_{-0.15}$ &$ 989\pm131$ &$ 173\pm51^{\dagger\dagger}$ &$ 5.7\pm1.7$ &$ >5.0$ &M \\
\hline \hline
\end{tabular}
\end{center}
\vspace{-0.4cm}
\tablecomments{
(1) Source ID. 
(2) Spectroscopic redshift.  
(3) Absolute UV luminosity, calculated from the F150W filter. 
(4) Dust extinction. 
(5) Average SFR over 10~Myr.  
(6) Stellar mass. 
(7) Fraction of the stellar mass produced by the young ($<10$~Myr) stellar population. 
(8) Rest-frame equivalent width of \oiii5008 line. 
(9) Rest-frame equivalent width of H$\beta$ line. 
(10) \oiii5008/H$\beta$ ratio. 
(11) \oiii5008/\oii3727 ratio. 
(12) NIRSpec observation mode (P: prism, M: Medium Resolution). For the sources observed in both modes, we adopt the results with the better S/N in the brightest emission line.
(4--7) values are estimated from the SED analysis with {\sc cigale}, while (8--11) values are derived from the NIRSpec observations. \\
$\dagger$ The \oii\ line wavelength falls in the detector gap. \\
$\dagger\dagger$ The H$\beta$ line wavelength falls in the detector gap, while we evaluate the H$\beta$ flux from the H$\gamma$ detection (see text). 
}
\end{table*}

\subsection{Line Flux Measurement}
\label{sec:flux}

We evaluate the fluxes of several key rest-optical emission lines for the $z_{\rm spec}$ sample. 
We focus on the line fluxes of \oiii5008, \oiii4960, H$\beta$, and \oii3728 in this paper. 
We perform the Gaussian + continuum profile fitting by fixing the redshift. We fit \oiii5008, \oiii4960, and H$\beta$ lines simultaneously, while we separately fit \oiii3727. 
In the fitting, we assume the same line width in the velocity frame among these emission lines. We also assume that the same underlying continuum among the \oiii5008, \oiii4960, and H$\beta$ lines and fix the line ratio of 3:1 between \oiii5008 and \oiii4960. We use the rest-frame 4100--5200~${\rm \AA}$ for the fit to obtain a stable result for the underlying continuum. 
For ID24, the wavelength of the H$\beta$ line falls in the detector gap, while the H$\gamma$ line is detected at S/N$\sim$3. We thus alternatively evaluate the line flux of H$\gamma$ and convert it to H$\beta$ with H$\gamma$/H$\beta$ of 0.471 for case B recombination with the electron temperature of 10$^{4}$~K without dust correction \citep{osterbrock1989} due to the negligible dust extinction estimated from our SED analysis (Section~\ref{sec:sed}).  

In Table~\ref{tab:prop}, we list the rest-frame equivalent width (EW) values of \oiii5008 and H$\beta$ and R3 ($\equiv$ \oiii5008/H$\beta$). When we do not securely constrain the H$\beta$ flux at more than S/N $=2$, we place 3$\sigma$ upper and lower limits for EW(H$\beta$) and R3, respectively. When the continuum is not well constrained (S/N $< 2$) from the fit, we estimate the underlying continuum by correcting the slit-loss for the line emission (see Section \ref{sec:slitloss}) and subtracting the line contribution to the F444W photometry. 
We confirm that our EW measurements are consistent with the independent analysis and measurement for MPT-ID3, 23, and 24 presented in \cite{tang2023} within the errors. 

We do not detect \oii\ with S/N $> 2$ from any of the $z_{\rm spec}$ sources. We thus place 3$\sigma$ upper limits on the \oii\ line and list the 3$\sigma$ lower limits of O32 ($\equiv$ \oiii5008/\oii3727) in Table \ref{tab:prop}. 
We do not apply the slit-loss corrections for this calculation. 
Due to the wavelength dependence of the \jwst\ point spread function, slit losses should be smaller for \oii\ compared to \oiii, and a slit loss correction would increase O32. Thus, the uncorrected lower limit for O32 is a conservative estimate. 

Note that there are several other line features that show  peak pixel values with S/N $\sim1-3$ at the expected wavelength positions (e.g., [Ne\,{\sc iii}], N\,{\sc v}) in the 1D spectra, while in most of the cases, the y-axis position of the positive signals within the extracted aperture is not exactly matched with that of the brightest \oiii5008 line. Although we cannot rule out the possibility of different spatial distributions for different emission lines\footnote{
For example, spatial offsets between the faint AGNs and the hosts have been reported in \cite{ding2022}, where highly ionized emission lines might be observed with spatial offsets. 
} or the effect of noise fluctuations in these low S/N spectra, we do not focus on the other emission lines in this paper. 

We also note that several spectra exhibit negative and weak positive traces distinct from the target positions (e.g., MPT-ID 7 and 23). We do not find any excessively bright background values that would indicate unknown open shutters in one or more nods. However, we observe small background variations in some observations among different nods, which likely cause the negative and positive traces due to over/under subtraction of the background. We further confirm that the relative positions of the negative and positive traces are consistent with cases where the background becomes relatively high in one nod.
The background subtraction and integration process using the standard 3-nod approach in the NIRSpec MSA observations\footnote{\url{https://jwst-docs.stsci.edu/jwst-near-infrared-spectrograph/nirspec-operations/nirspec-dithers-and-nods/nirspec-mos-dither-and-nod-patterns}} ensures that variations in the background do not impact shutters containing the target, and thus our flux estimate remains unaffected by this effect.

\subsection{Slitloss Correction}
\label{sec:slitloss}

As mentioned in Section~\ref{sec:NIRSpec_reduction}, we do not apply the default slitloss correction implemented in the calibration pipeline.
Instead, we evaluate the scaling factor for the slit-loss correction in the two following approaches. 
In the first approach, we convolve and integrate the 1D spectrum with the NIRCam F444W filter response function. By comparing the resulting value with the actual total galaxy flux measured with the NIRCam F444W filter, we can assess the flux loss in the 1D spectrum and derive a scaling factor.
In the second approach, we calculate the flux in the F444W image enclosed within the $0\farcs2\times0\farcs46$ shutter based on the source and MSA shutter positions and estimate the scaling factor to match it with the total flux measurement. 
We find that MPT-ID4 and ID7 show negative scaling factors in the first approach, indicating that the integrated F444W spectral fluxes are dominated by noise fluctuation, even with the inclusion of the strong line emission in these sources. 
We thus adopt the second approach for MPT-ID4 and ID7, while we use the first for the rest of our targets. 
When we compare these two approaches using some bright targets, we find that the scaling factor can be different by $\sim$10--40\%. This may be due to unknown offsets of targets within the MSA shutter related to the accuracy of the MSA alignment or may indicate uncertain relative photometric calibration between NIRCam and NIRSpec, which cannot be corrected in the second approach. 
Diffraction losses within NIRSpec \citep{Jakobsen2022}, which are not included in the second approach, could also be a possible reason for the difference. 
An extended \oiii+H$\beta$ structure has been identified around a lensed galaxy at $z=8.5$ \citep{fujimoto2022c}, and the differential spatial distribution between the ionized gas and the underlying continuum emission is also another unknown uncertainty in both approaches. We thus caution the readers that additional uncertainty might exist in the measurements due to the slit-loss correction steps.

\subsection{SED Analysis}
\label{sec:sed}

We analyze the physical properties of the seven spectroscopically confirmed NIRCam galaxies 
using {\sc cigale}. 
The details will be presented in Burgarella et al.\ (in prep.), and here we briefly explain our SED model. 
We include the \hst+\jwst/NIRCam photometry as well as the EW(\oiii5008) in the fit. 
We use the stellar synthesis population model of BPASS-v2 \citep{eldridge2017} with the lowest available $Z\_{\rm BPASS} = 0.0001$, where $Z_{\rm \odot}$ = 0.014. 
We assume a delayed + final exponential burst (younger than 10~Myr) star-formation history (SFH): SFR($t$) $\propto t/\tau_{0}^{2}\,\times\,$exp$(-t/\tau_{0})$ + $k\,\times\,$exp($-t/\tau_{1}$) with the mass fraction from the recent burst activity of $f_{\rm burst}=0.0-0.5$. 
We adopt the dust attenuation law from \cite{calzetti2000} for the stellar continuum and a screen model with an SMC extinction curve \citep{pei1992} for the nebular emission (continuum + lines). During the SED fitting, we assume the $E(B-V)_{\rm stellar} = 0.5 \times E(B-V)_{\rm line}$. 
Interested readers are referred to \cite{boquien2019} for specific modeling procedures using {\sc cigale}. 
The results for SFR, $M_{\rm star}$, $E(B-V)$, $f_{\rm burst}$ are summarized in Table~\ref{tab:prop}. From the median values, our NIRCam-selected spec-$z$ confirmed sources are generally characterized as SFR $\simeq\,4\,M_{\odot}$~yr$^{-1}$, $M_{\rm star}\simeq10^{8}\,M_{\odot}$~yr$^{-1}$, $E(B-V)\simeq0.1$, and $f_{\rm burst}\simeq$~10\%. 
Note that the SED analysis with the spatially-integrated photometry may systematically underestimate the stellar mass, especially in the strong emission line systems \citep{clara2022b, papovich2022}. 
In our analysis, this effect is mitigated by constraining the $f_{\rm burst}$ factor to the range 0--50\%, which provides the median of $f_{\rm burst}\simeq20$\% in our sample. 
The $f_{\rm burst}\simeq10$\% is 
consistent with recent results of the pixel-based spatially-resolved SED analysis for lensed galaxies at $z=5-8.5$ with NIRCam \citep{clara2022b}. 
The consistency and potential differences among different SED models are further discussed in Burgarella et al. (in prep.).

\section{Results \& Discussions}
\label{sec:result}

\subsection{Validity of High-redshift Galaxy Selection}
\label{sec:zspec}

\begin{figure*}[t]
\begin{center}
\includegraphics[trim=0.2cm 0cm 0cm 0cm, clip, angle=0,width=0.85\textwidth]{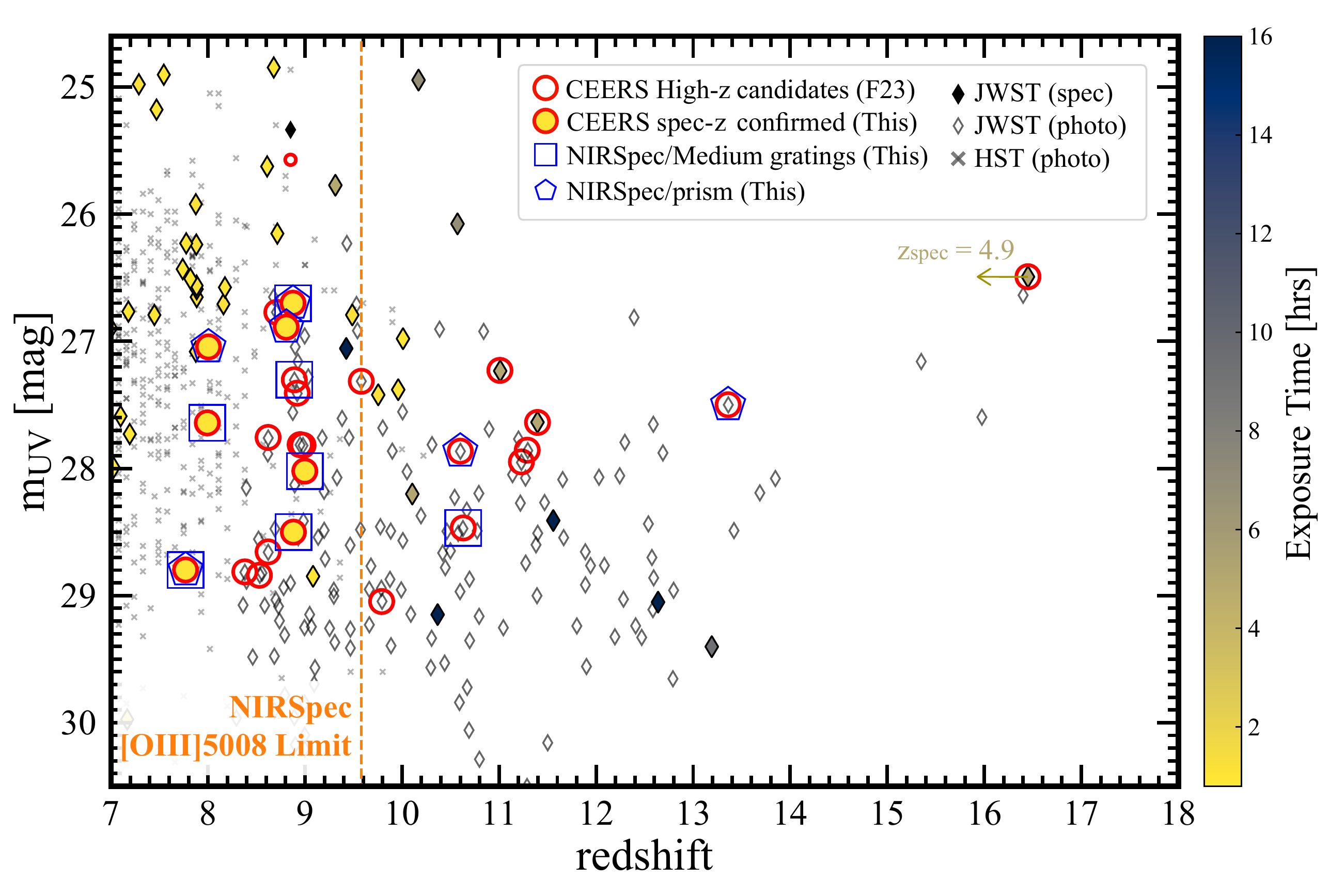}
\end{center}
\vspace{-0.4cm}
 \caption{
Status of spectroscopic confirmation at $z\gtrsim8$ with \jwst. 
The vertical axis represents the apparent UV magnitude. 
The red open circles show the 26 candidates at $z\gtrsim8.5$ in CEERS presented in \cite{finkelstein2023} (F23), and the filled circles indicate the sources spectroscopically confirmed in this study. 
The open blue pentagons and squares indicate the observation mode of prism and Medium gratings, respectively. 
Among the 26 candidates, the highest-redshift candidate at $z_{\rm phot}\sim17$ is spectroscopically determined to have $z=4.9$, 
while two other candidates are confirmed at $z=11.4$ and $z=11.0$ \citep{arrabal-halo2023a, harikane2023b}. 
The color diamonds indicate other $z_{\rm spec}$ confirmed galaxies taken from recent \jwst\ studies  
from lensing cluster fields \citep{williams2022, roberts-borsani2022b,wang2022, mascia2023} and general fields \citep{tang2023, curtis-lake2022, bunker2023, arrabal-halo2023a, arrabal-halo2023b, harikane2023b}. 
No magnification corrections are applied to the lensed sources. 
The color of the circles and diamonds corresponds to the exposure time of the \jwst/NIRSpec denoted in the color bar. 
The grey crosses and open diamonds present photometric high-$z$ candidates from the literature identified with \hst\ and \jwst, respectively. 
The dashed vertical line presents the redshift limitation to detect the \oiii5008 line with NIRSpec, indicating the potential challenges of the $z_{\rm spec}$ determination beyond $z\sim9.6$ with a short exposure for faint objects due to the lack of bright emission lines. 
\label{fig:zspec}}
\end{figure*}

\begin{figure}
\includegraphics[trim=0.2cm 0cm 0.2cm 0cm, clip, angle=0,width=0.47\textwidth]{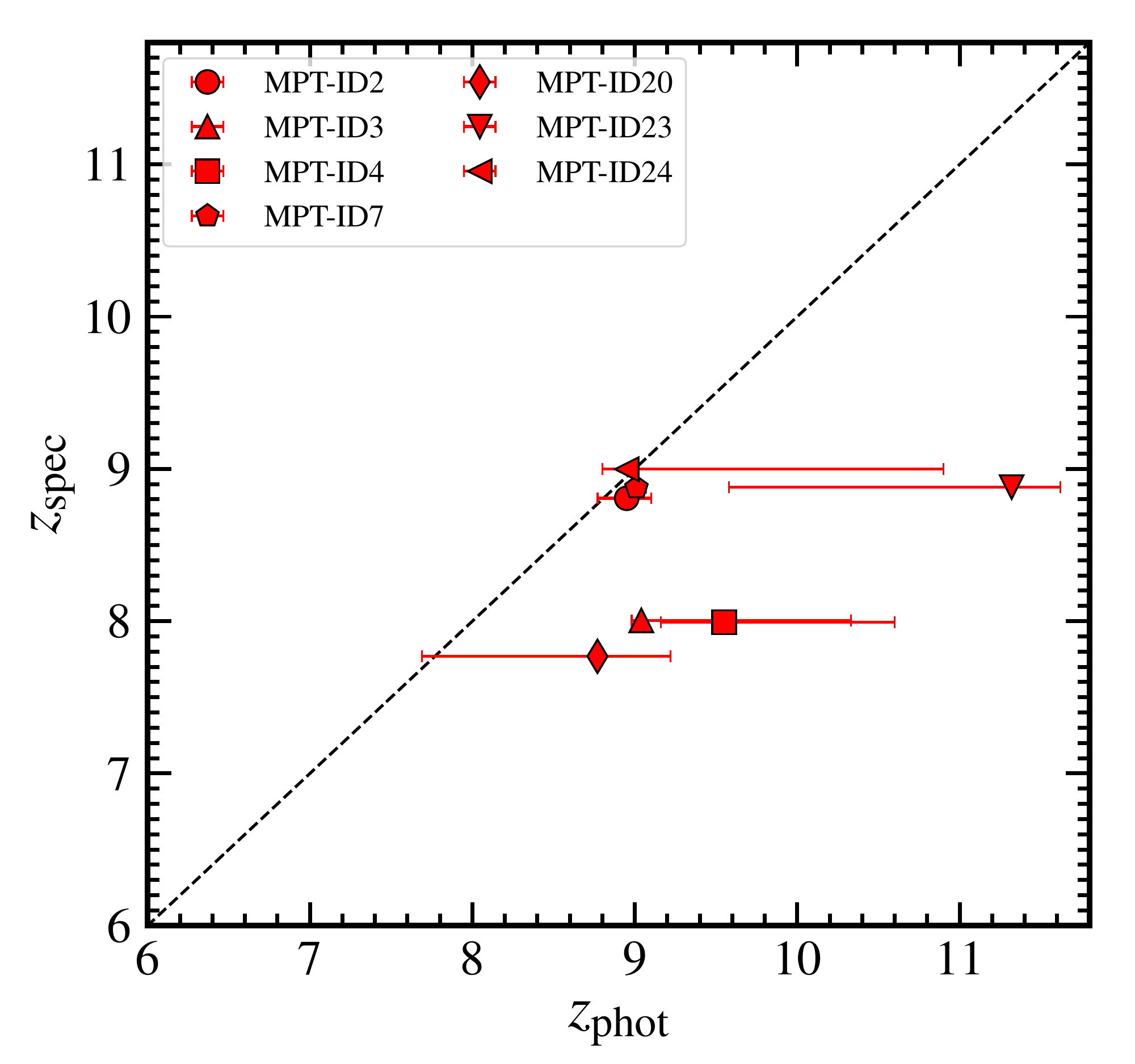}
 \caption{
Photometric vs. spectroscopic redshift for the NIRCam-selected galaxies with $z_{\rm phot}\gtrsim8.5$. 
Six out of seven sources show $z_{\rm spec} < z_{\rm phot}$, though all are consistent within 2$\sigma$. 
\label{fig:zcomp}}
\end{figure}

In Figure~\ref{fig:zspec}, we present the distribution of UV apparent magnitudes ($m_{\rm UV}$) versus redshift (spectroscopic where available, otherwise photometric) for NIRCam-selected galaxies from CEERS \citep{finkelstein2023} and other recent \jwst\ photometric \citep{bouwens2021b, naidu2022b, castellano2022, adams2022, atek2022, donnan2023, donnan2023b, harikane2023, bouwens2022c, bouwens2022d, morishita2022, bradley2022, perez-gonzalez2023} and spectroscopic studies \citep{arrabal-halo2023a, arrabal-halo2023b, curtis-lake2022, williams2022, roberts-borsani2022b, wang2022, mascia2023, tang2023, bunker2023, harikane2023}. 
When necessary, we convert the absolute UV magnitudes reported in the literature to the apparent magnitudes based on their $z_{\rm spec}$ or $z_{\rm phot}$ values. 
For the $z_{\rm spec}$ sources, the color of the symbols corresponds to the exposure time of the follow-up NIRSpec observations. No magnification correction is applied to the lensed sources. 

Among our 11 NIRCam-selected high-redshift galaxy candidates, we confirm the spectroscopic redshifts of 7 galaxies at $z=7.769-8.998$. 
Our CEERS observations demonstrate the capability of NIRSpec to determine the redshift even with a $\sim50$-min exposure down to $m_{\rm UV}\simeq29$~mag at $z\simeq9$  ($M_{\rm UV}\simeq-18$ mag). 
On the other hand, our results suggest that the $z_{\rm spec}$ confirmation at $z\gtrsim10$ is still challenging due to the lack of strong emission lines observable at these higher redshifts within the NIRSpec wavelength coverage. 
The successful cases at $z\gtrsim10$ in the literature are achieved mainly by the identification of the Lyman-$\alpha$ break \citep[e.g.,][]{curtis-lake2022, arrabal-halo2023a, arrabal-halo2023b}, which is likely owing to deep exposures ($>5$~hrs) and/or the brightness of the target ($m_{\rm UV}<27.5$~mag) even with the similarly short exposure ($\sim1$~hr). 
The slitloss may also affect the success ratio of the $z_{\rm spec}$ determination for a given source flux. 

If we examine the success ratio based on each redshift range according to 
the original $z_{\rm phot}$ estimates of our targets, we achieve the spectroscopic confirmation in six out of seven ($\simeq90\%$) targets with $8.8\lesssim z_{\rm phot}\lesssim9.5$. 
Owing to the homogeneous sample selection and sensitivity in the follow-up spectroscopy, our results indicate that the classical 
$z_{\rm phot}$ technique assures at least $\sim90\%$ of the high-redshift galaxy selection at $z\simeq9$. 
This is helpful to interpret the recent and future UVLF results out to $z\simeq9$ that are also measured via similar dropout or $z_{\rm phot}$ techniques with NIRCam \citep[e.g.,][]{donnan2023, harikane2023, bouwens2022c, bouwens2022d}.  
Importantly, CEERS does not include a NIRCam-based dropout filter for the $z\simeq9$ galaxy selection (e.g., F090W), and instead, it relies on the \hst/ACS F814W filter with a 5$\sigma$ limiting AB magnitude for a point source of 28.3~mag \citep{finkelstein2023}. 
Therefore reliable high-redshift galaxy selection could be further improved by the addition of a deeper NIRCam dropout filter\footnote{
For example, the GLASS-JWST \citep{treu2022} survey includes the F090W filter, where robust $z=11-13$ candidates have been reported \citep[e.g.,][]{castellano2022, naidu2022a}.
}.

Interestingly, the majority of our $z_{\rm spec}$ sample has a $z_{\rm spec}$ value smaller than the $P(z)$ peak. In Figure~\ref{fig:zcomp}, we compare the $z_{\rm phot}$ and $z_{\rm spec}$ values for our $z_{\rm spec}$ sample. Although all $z_{\rm spec}$ values fall within the 95\% confidence range of $P(z)$ (see Figure~\ref{fig:spectra}), three out of seven sources show $z_{\rm spec} < z_{\rm phot}$ beyond the 68\% confidence interval. 
The $z_{\rm phot}$ value and the entire $P(z)$ shape largely depend on the galaxy models used in the SED fitting. 
In our original target selection (Section \ref{sec:NIRCam_data}), the high-redshift galaxy candidates are selected by using the SED fitting code {\sc eazy} \citep{brammer2008} based on the FSPS-based \citep{conroy2010} default 12 template sets in {\sc eazy} and 6 additional templates presented in \cite{larson2022b}. These six additional templates are created by combining stellar population spectra from BPASS \citep{eldridge2009} and {\sc Cloudy} \citep{ferland2017} to recover the rest-frame UV blue color space of young galaxies, which significantly recovers the color space of simulated early galaxies in a semi-analytic galaxy model \citep{yung2022}. 
However, the high fraction of $z_{\rm spec} < z_{\rm phot}$ observed in our sample might indicate that the SED templates generally used in the high-redshift galaxy studies are still insufficient to recover the rest-frame UV color space of $z\gtrsim8.5$ galaxies, or that the modeling of the IGM is incomplete in some way (though likely \emph{not} due to the damping wing effect discussed in \citealt{curtis-lake2022}, which would lead to redshift differences much smaller than we observe here).  
This could also show the possible limitations inherent in using a single photo-$z$ code \citep[e.g.,][]{dahlen2013}. 
Another reason might be the difficulty of determining the redshift at $z\gtrsim9.5$ due to the lack of strong emission lines in the NIRSpec wavelength coverage. Then, while the true redshifts are distributed on both sides of the peak of $P(z)$ according to uncertainties in the broadband photometry, the spec-$z$ confirmation may be biased to $z\lesssim9.5$ when the sensitivity is not deep enough to capture the Lyman-$\alpha$ break \citep{curtis-lake2022, robertson2022, roberts-borsani2022b} and has to rely on the line identification to determine $z_{\rm spec}$. 
The absence of the NIRCam-based dropout filter may also at least partially contribute to the $>$1$\sigma$ difference between $z_{\rm phot}$ and $z_{\rm spec}$ for these sources.
For further discussion on this topic, readers are referred to \cite{arrabal-halo2023b} (see~Section 4.2).

Of the other four targets with $z_{\rm phot}>10$, one target turns out to have $z_{\rm spec}=8.8802$ (MPT-ID 23), while no robust lines are identified in the other three targets. 
The lack of lines might be because \oiii+H$\beta$ is redshifted beyond the NIRSpec wavelength range at $z\gtrsim10$. 
Given that one out of four has been confirmed at $z_{\rm spec}<10$, this would imply that 75\% (=3/4) candidates might be truly at $z>10$, although we cannot rule out the alternative possibility that the remaining three candidates have entirely incorrect redshifts at $z \ll 10$, and their emission lines are fainter than the current NIRSpec sensitivity. 
Once the success rate of the high-redshift galaxy selection is also confirmed at $z_{\rm phot}\gtrsim10$, 
again, it will be widely helpful to interpret the photometric-based UVLF measurements, which are now explored out to $z\sim17$ with NIRCam \citep[e.g.,][]{harikane2023, donnan2023, donnan2023b, bouwens2022c, bouwens2022d, finkelstein2023}. 
This will enable us to verify the recent arguments of the high abundance of bright galaxies at these early epochs of the universe, which may challenge the current galaxy formation and evolution models.

\subsection{Properties of Galaxies at $z=$ 8--9}
\label{sec:prop}

The high $z_{\rm spec}$ confirmation rate ($\simeq90$\%) for the $z_{\rm phot}=8.5-9.6$ targets from the homogeneous LBG selection
indicates that our spec-$z$ sample is representative of the majority of the NIRCam-selected LBG population at this redshift. 
Moreover, recent UVLF studies suggest that the characteristic UV luminosity is $M_{\rm UV}^{\star}\sim-21$~mag in the Schechter function form \citep[e.g.,][]{finkelstein2015, bouwens2021, harikane2023}.  
Therefore, our NIRCam-selected $z_{\rm spec}$ sources, ranging in the UV luminosity range $M_{\rm UV}\in[-21,-18]$, are sub-$L^{\star}$ populations down to $\simeq0.06\times L^{\star}$, which is helpful to understand the typical physical properties of the abundant faint population without gravitational lensing uncertainties. 
In the following subsections, 
we thus carry out the first census of the physical properties of spectroscopically-confirmed sub-$L^{\star}$ galaxies at this redshift range and investigate the potential difference from bright galaxies ($M_{\rm UV}\simeq-22$) identified in previous \hst\ studies and also spectroscopically confirmed at similar redshifts (Larson et al., in prep.).

\subsubsection{SFR vs. $M_{\rm star}$}
\label{sec:sfr-mstar}

\begin{figure}
\begin{center}
\includegraphics[trim=0cm 0cm 0cm 0cm, clip, angle=0,width=0.47\textwidth]{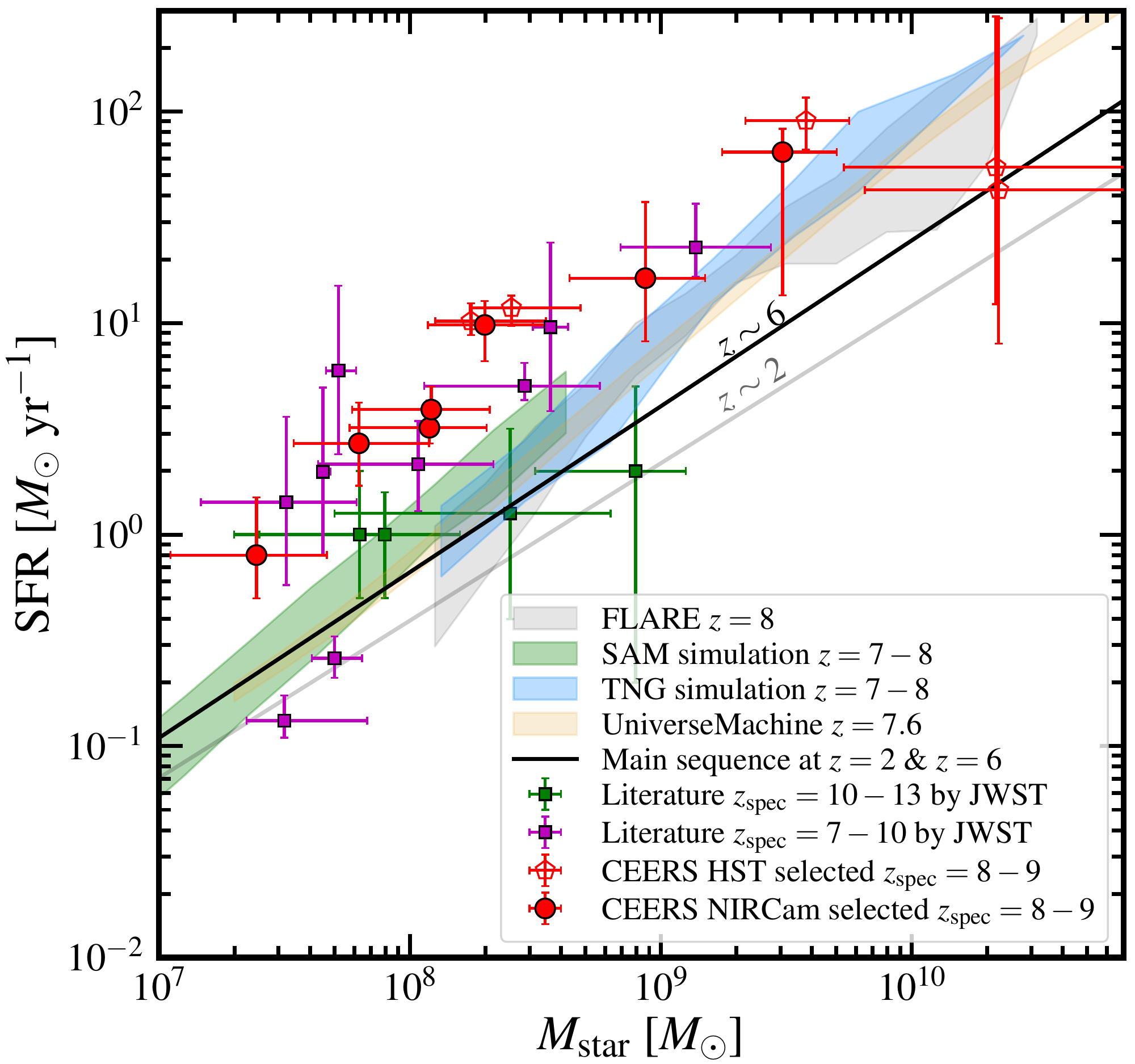}
 \caption{
SFR vs. $M_{\rm star}$. 
The filled and open symbols represent the spec-$z$ and photometric samples, respectively.  
The filled red circles show our $z_{\rm spec}$ sample. 
The magenta and green squares are recent \jwst\ results for lensed  galaxies at $z\sim7-10$ \citep{tacchella2022, williams2022, roberts-borsani2022b, heintz2022b} and for general field galaxies at $z\sim10-13$ \citep{robertson2022, curtis-lake2022}, respectively. 
The red pentagons show recent \hst+IRAC results for bright galaxies at $z\sim8-9$ in the EGS field \citep{tacchella2022a}.  
The black and grey lines denote the best-fit relations at $z=2$ and $z=6$ \citep{iyer2018}. 
The color-shaded regions present the 16-84th percentile of simulated galaxies \citep{yung2019b, yung2022, behroozi2019, behroozi2020, wilkins2022a, wilkins2022b}. 
Our NIRCam-selected $z_{\rm spec}$ galaxies show high sSFR ($\simeq40$~Gyr$^{-1}$) compared to the $z=2-6$ main sequence. 
\label{fig:sfr-ms}} 
\end{center}
\end{figure}

Figure~\ref{fig:sfr-ms} presents the SFR--$M_{\rm star}$ relation. 
For comparison, we show the main sequence estimated at $z=2$ and $z=6$ \citep{iyer2018},
recent \jwst\ results for the sources spectroscopically confirmed at $z=7-13$ \citep{tacchella2022, williams2022, roberts-borsani2022b, heintz2022b, robertson2022, curtis-lake2022}, 
the \hst-selected, brighter LBGs ($M_{\rm UV}\simeq-21.5$) spectrosopically confirmed at $z=8-9$ in CEERS (Larson et al. in prep.), and the 16-84th percentile of the distribution of simulated galaxies at $z=7-8$ \citep{yung2019b, yung2022, behroozi2019, behroozi2020, nelson2019, wilkins2022a, wilkins2022b}. 
Note that we use the SED-based SFR estimate (Section \ref{sec:sed}), instead of using the H$\beta$ line. This is because of the potential uncertainty in the slit-loss correction (Section \ref{sec:slitloss}). Also, an extended \oiii+H$\beta$ structure has been recently reported around a lensed galaxy at $z=8.5$, which might make the H$\beta$-based SFR measurement to be overestimated by counting the H$\beta$ emission caused by other physical mechanisms such as powerful outflows \citep{fujimoto2022c}.  

In Figure~\ref{fig:sfr-ms}, we find that our NIRCam-selected sample (red filled circles) has a SFR--$M_{\rm star}$ relation generally consistent with the recent \jwst\ results for lensed galaxies at $z=7$--10 (magenta squares) down to $\sim7\times10^{7}\, M_{\odot}$. 
These galaxies from our and recent \jwst\ studies are generally less star-forming and less massive than the \hst-selected bright galaxies at similar redshifts (red open pentagons). 
We also find that most of our $z_{\rm spec}$ sample and the lensed galaxies in the literature fall above the $z=6$ main sequence by $\sim0.5$~dex scale beyond the errors at a given $M_{\rm star}$, showing a relatively high specific SFR (sSFR $\equiv$ SFR/$M_{\rm star}$) of $\simeq40$~Gyr$^{-1}$. 
In comparison, two out of five CEERS \hst-selected galaxies fall close to the $z=6$ main sequence.
This finding is in line with recent \jwst\ results that less luminous LBGs at $z\sim7-8$ have similarly high sSFR of $\simeq80$~Gyr$^{-1}$ \citep{endsley2022b}. 
As discussed in \cite{endsley2022b}, the high sSFR in the less luminous LBGs, highlighted by our $z_{\rm spec}$ sample and the lensed galaxies, is likely caused by young stellar populations.  
This is consistent with our SED results showing the presence of the young stellar populations ($<10$~Myrs) accounting for the total mass of $\simeq20$\% of the galaxy mass (Table~\ref{tab:prop}). 
The increasing sSFR trend as a function of redshift has been observed at lower redshifts \citep[e.g.,][]{tacconi2013, speagle2014, tasca2015, khusanova2020}. Our results may indicate that this increasing trend continues at least out to $z=8-9$, 
which is predicted from the simulations due to the increased gas accretion rate onto dark matter halos at higher redshifts \citep[e.g.,][]{behroozi2013}. In fact, we also confirm the general agreement with predictions from the simulations within the errors. 

Note that a different SFH assumption may impact our SED estimates. 
For example, our parametric (delayed + final burst $<$ 10~Myr) approach might miss an extended tail of older star-forming activity in the SFH. However, we confirm that the mass estimate increases by a factor of $\sim$2 at most in our sample by using the \texttt{Dense Basis} \citep{iyer2019} fitting code with a non-parametric SFH assumption, which is still insufficient to match the distribution with the main sequence at $z=2-6$.  
The general agreement with measurements from a variety of literature sources for lensed galaxies also suggests that the impact from different SFH assumptions and SED modelings is likely small. 
It should also be noted that the slight excess of the observational results compared to the distribution of the simulated galaxies at $z=7-8$ might be caused by the observation bias, which is not counted in the simulated galaxies due to the difficulty of quantifying the spec-$z$ confirmation process in the observation. However, the main sequence at $z=2-6$ is also measured by observations that should include a similar bias. 
Therefore, the relatively high sSFR of the $z=8-9$ galaxies, compared to that of the $z=2-6$ galaxies, is unlikely explained by the observation bias. 

\subsubsection{\oiii+H$\beta$ EW}
\label{sec:oiii_ew}

\begin{figure}[t]
\includegraphics[trim=0cm 0cm 0cm 0cm, clip, angle=0,width=0.47\textwidth]{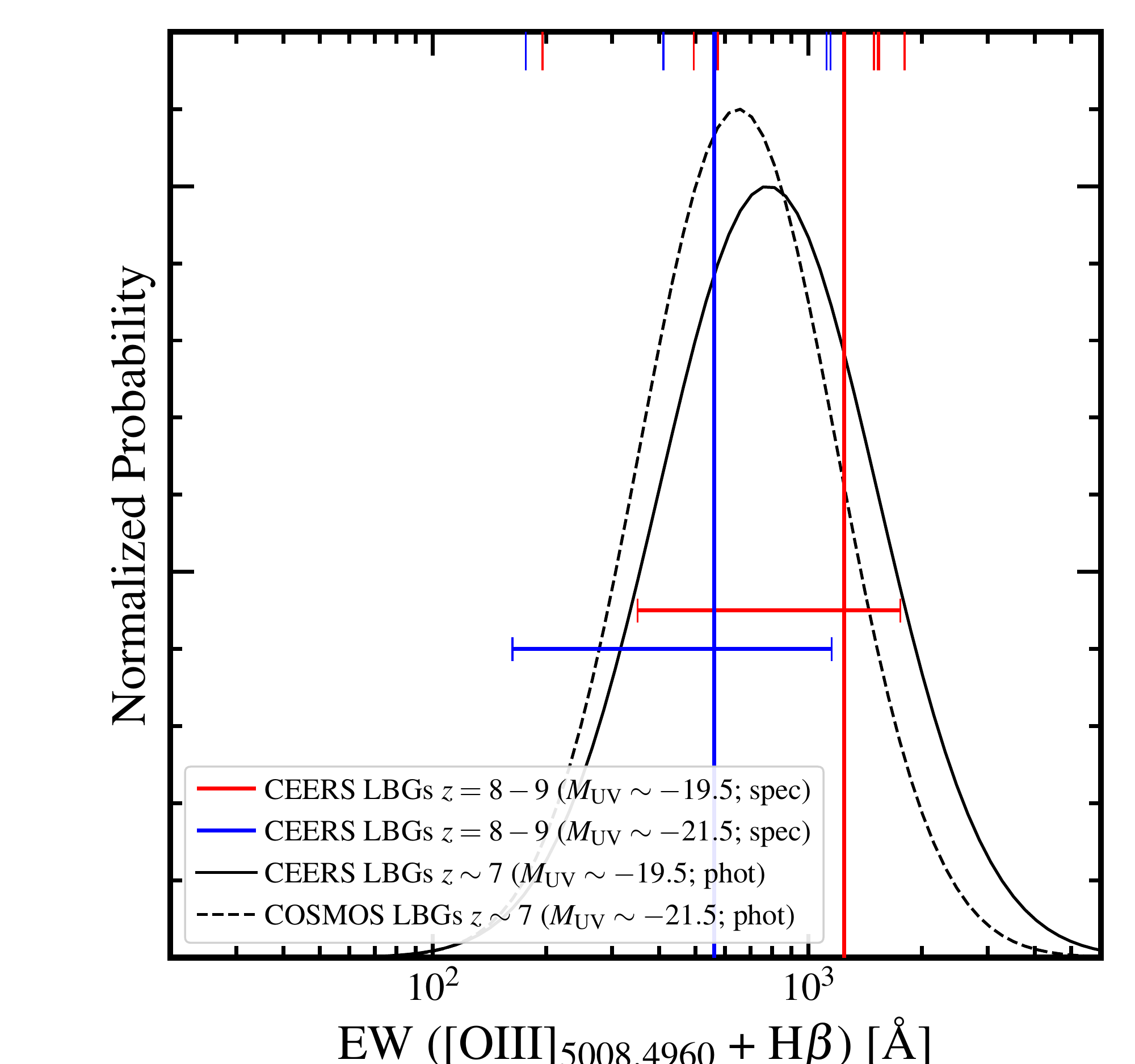}
 \caption{
 Rest-frame EW distribution of \oiii+H$\beta$. 
 The red and blue lines and error bars  present the median and 16-84th percentiles of fainter (NIRCam-selected) and brighter (\hst-selected, Larson et al., in prep.) LBGs with spectroscopic redshifts $z=8-9$, where the red and blue vertical bars at the top show the individual measurements. 
 The black solid and dashed 
 curves show the best-fit log-normal distribution estimated from SED analysis for photometric samples of 36 luminous LBGs at $z\simeq6.6-6.9$ identified in the COSMOS field ($M_{\rm UV}\sim21.5$~mag) and 118 less luminous LBGs at $z\sim6.5$--8 in CEERS \citep{endsley2022b}. 
\label{fig:ew}}
\end{figure}

In the previous subsections, the sSFR values determined for our $z_{\rm spec}$ sources are higher than those for $z\sim6$ main-sequence galaxies, which is likely because of the young stellar populations at higher redshifts. 
From photometric-based SED analysis, high EW(\oiii+H$\beta$) values ($>1000 {\rm \AA}$) have been inferred for galaxies at $z\simeq$ 6--9 in previous \hst+IRAC studies \citep[e.g.,][]{labbe2013, debarros2019, endsley2021a}, where the high EW(\oiii+H$\beta$) is thought to be caused by the light from very young (1--10~Myr) stellar populations \citep[e.g.,][]{amorin2017, tang2019, vanzella2020}. To have an independent insight into the stellar population and ISM properties of our galaxies, we thus compare the EW(\oiii+H$\beta$) values of our NIRCam-selected galaxies with other galaxy populations in this subsection.  

In Figure~\ref{fig:ew}, we show the median (red line) and 16-84th percentile (red bar) of the rest-frame \oiii+H$\beta$ EW measurement for our $z_{\rm spec}$ sample. In the same manner, we also measure and show the median and 16--84th percentile of \oiii+H$\beta$ EW values for a sample of brighter \hst-selected LBGs in CEERS (blue line and bar), also spectroscopically confirmed at $z=8-9$ (Larson et al., in prep). 
The 16--84th percentile is estimated by assuming a log-normal probability distribution for the EW measurements of each source and summing them to define the entire probability distribution. The upper limits of H$\beta$ EW (MPT-ID2, 4) are also included in this calculation, assuming that zero and the 3$\sigma$ error correspond to its center and standard deviation of the log-normal distribution.
 For comparison, we also present the EW distribution estimated from the photometric-based SED analysis for LBGs at $z\sim7$ with similar UV luminosity ($M_{\rm UV}\sim-19.5$; solid line) with \jwst\ data and brighter LBGs ($M_{\rm UV}\sim-21.5$; dashed line) at $z\sim7$ \citep{endsley2022b}. 

We obtain median EW(\oiii+H$\beta$) values of $1100_{-730}^{+560} {\rm \AA}$ and $570_{-410}^{+590} {\rm \AA}$ for the spec-$z$ confirmed, NIRCam-selected and \hst-selected luminous galaxies, respectively. 
These measurements are generally much higher than $z=1-4$ galaxies with similar SFR ($\simeq100-200{\rm \AA}$; \citealt{reddy2018b}), but consistent with the previous photometric-based measurements for $z=7-8$ galaxies ($\simeq600-800{\rm \AA}$; e.g., \citealt{labbe2013, debarros2019, endsley2021a}). 
We find that less luminous galaxies have higher EW(\oiii+H$\beta$) values, while the difference is still consistent within the $1\sigma$ ranges. 
The slight increase of EW(\oiii+H$\beta$) in less luminous galaxies agrees with the trend observed in the recent photometric-based results (solid and dashed curves). 
The increasing trend of EW(\oiii+H$\beta$) with decreasing metallicity has been observed at $z=1-4$ \citep{reddy2018b}. However, in extremely metal-poor systems, the \oiii\ emission should be suppressed due to the low oxygen abundance, and the EW(\oiii+H$\beta$) value should be significantly smaller for a given value of the ionization parameter.  
Therefore, the slight increase of EW(\oiii+H$\beta$) observed in our NIRCam-selected galaxies at $z=8-9$ would indicate that these less luminous galaxies are more metal-poor systems than the luminous galaxies, but not extremely metal-poor systems. 
The high EW(\oiii+H$\beta$) value might also indicate a higher ionization parameter. 
Recent studies report the detection of the intense nebular emission from highly ionized carbon in the rest-frame UV from $z>7$ galaxies \citep{stark2015, stark2017, laporte2017b, mainali2018, schmidt2017, hutchison2019}. From a deep IRAC 5.8$\mu$m band stacking for $z\sim8$ LBGs, \cite{stefanon2022} report a large H$\alpha$+\nii\ EW and subsequently derive a very high ionizing photon production efficiency of $\log$(\xion/Hz~erg$^{-1}$) $=25.97^{+0.18}_{-0.28}$. 
The redshift evolution of the electron density has also been reported, reaching $n_{\rm e}\gtrsim300$~cm$^{-3}$ at $z\sim9$ \citep{isobe2023}. 
These results suggest that the ISM in higher redshift, younger galaxies is more highly ionized than in lower-$z$ galaxies, which may lead to maintaining the high EW(\oiii+H$\beta$) distribution even in our NIRCam-selected less luminous galaxies at $z=8-9$.   
We caution that our results might be biased toward the high EW systems, given the fact that the spec-$z$ of high EW systems is easily confirmed via bright emission lines, although our $z_{\rm spec}$ sample represents $\simeq$ 90\% of the less luminous LBGs at $z_{\rm phot}=8.6-9.6$ from the homogeneous selection (Section \ref{sec:zspec}). 
The presence or absence of the increasing trend of EW(\oiii+H$\beta$) distribution as a function of UV luminosity and redshift will be further investigated with larger spectroscopic samples in a separate paper. 

\subsubsection{Ionizing photon production efficiency}
\label{sec:xi_ion}

\begin{figure}[t]
\begin{center}
\includegraphics[trim=0cm 0cm 0cm 0cm, clip, angle=0,width=0.5\textwidth]{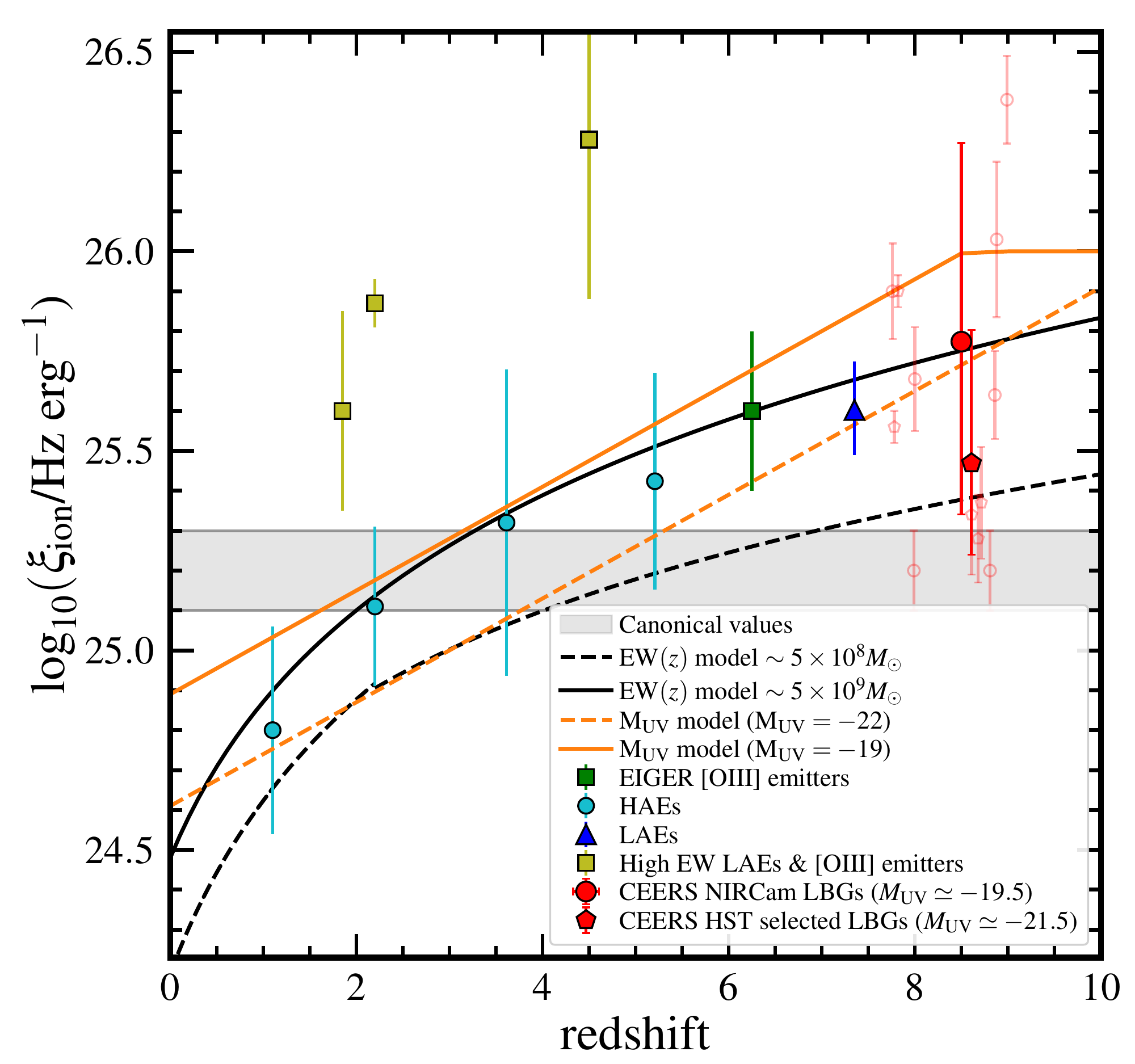}
 \caption{
Redshift evolution of the ionizing photon production efficiency $\xi_{\rm ion}$. 
The red-filled circle and pentagon with error bars present the median and 16-84th percentiles of the spec-$z$ confirmed, less luminous and luminous LBGs at $z=8-9$ selected by NIRCam and \hst\ (Larson et al.\ in prep.), respectively, 
where the small, light red open symbols show the individual measurements. 
Other color symbols are taken or calculated from the literature based on spectroscopy, showing the median/average values for a variety of populations: \oiii\ emitters at $z\sim6$ (green squares: \citealt{matthee2022}), 
Lyman $\alpha$ emitters at $z\sim7$ (LAEs, blue triangle; \citealt{stark2015, stark2017}), H$\alpha$ emitters at $z\sim1-2$ (HAEs, cyan circles; \citealt{matthee2017, atek2022, gonzalo2023}), and LAEs and \oiii\ emitters with high EWs (deep yellow squares; \citealt{tang2019, maseda2020, naidu2022a}).  
The grey shade shows the canonical value \citep[e.g.,][]{robertson2013}, and  
the dashed and solid curves are predictions from empirical models based on H$\alpha$ EW (black curves; \citealt{faisst2016, matthee2017}) and UV luminosity (orange curve; \citealt{finkelstein2019}). 
\label{fig:xi_ion}}
\end{center}
\end{figure}

To place our $z_{\rm spec}$ sample in the context of cosmic reionization, we also estimate the production efficiency of ionizing photons, $\xi_{\rm ion}$. We measure the $\xi_{\rm ion}$ values via the Balmer recombination line approach \citep[e.g.,][]{bouwens2016b} based on the H$\beta$ line in the same manner as \cite{matthee2022} by assuming a zero escape fraction of ionizing photons. 
A correction for dust attenuation is applied based on our SED fitting results (Section \ref{sec:sed}). For comparison, we measure $\xi_{\rm ion}$ also for the $z_{\rm spec}$ confirmed \hst-selected, luminous LBGs (Larson et al., in prep.) in the same manner. 

In Figure~\ref{fig:xi_ion}, we show our \xion\ estimates with recent spectroscopic results in different galaxy populations at high redshifts \citep{stark2015, stark2017, matthee2017, matthee2022, atek2022, naidu2022a, tang2019, maseda2020, gonzalo2023} 
and the canonical value that has been used when modeling the contributions of galaxies to the cosmic reionization \citep[e.g.,][]{robertson2013}. 
To perform a statistical comparison, we present the median and 16-84th percentile as the error bar for our \xion\ estimates of our $z_{\rm spec}$ sample by assuming a log-normal probability distribution for the \xion\ measurement of each source. Upper limits for H$\beta$ line fluxes (MPT-ID2, 4) are also included by assuming that the canonical value and $\pm0.1$ correspond to its center and standard deviation of the log-normal distribution, respectively. We obtain $\log$(\xion/Hz~erg$^{-1}$) $=25.77^{+0.50}_{-0.43}$ and $25.47^{+0.33}_{-0.23}$ for the NIRCam- and \hst-selected sources, respectively. 
We find that these median \xion values in our sample are much higher than the measurements at $z\sim1$--2 and the canonical value, 
but consistent with the recent result for the \oiii\ emitters (green square) and Lyman-$\alpha$ emitters (blue triangle) at $z\sim6-7$, and the strong line-emitting galaxies at $z\sim2-4$ (yellow squares). 
Our results are also consistent with the deep IRAC 5.8$\mu$m band stacking results for $z\sim8$ LBGs (25.97$^{+0.18}_{-0.28}$; \citealt{stefanon2022}).
We also find that less luminous, NIRCam-selected galaxies have higher $\xi_{\rm ion}$ than that of \hst-selected galaxies. 
Although these two measurements are consistent within the 1$\sigma$ ranges, 
this trend also aligns with the empirical redshift evolution models, which show increasing \xion\ with increasing mass (black curves; \citealt{faisst2016, matthee2017}) and UV luminosity (orange curves; \citep{finkelstein2019}), suggesting that faint galaxies play a key role in cosmic reionization.

As discussed in previous studies \citep[e.g.,][]{finkelstein2019, stefanon2022, matthee2022}, the \xion\ estimates higher than the canonical values require the modest ($\ll20\%$) escape fractions of ionizing photons \citep[e.g.,][]{davies2021} or that the contribution from faint galaxies ($M_{\rm UV}>-17$; e.g., \citealt{matthee2022b}) is minor. Albeit the small statistics, our $z_{\rm spec}$ sample represents the majority ($\simeq90\%$) of the LBGs at $z_{\rm phot}=8.6-9.6$ (Section \ref{sec:zspec}), which may support the arguments of the modest escape fraction or the negligible contribution from further faint galaxies. 
 However, the high $\xi_{\rm ion}$ measurements observed in galaxies $z\gtrsim6$ are challenging to the current physics-based galaxy formation models \citep[e.g.,][]{wilkins2016, yung2020a} with standard model components (e.g., power-law IMFs and simple stellar population SEDs). A top-heavy IMF and metal-free star SED in galaxies at these epochs are potential explanations to account for this discrepancy \citep[e.g.,][]{zackrisson2011, trussler2022}. Together with the faint-end measurements of the UVLFs at $z>8$, the contribution of the faint LBG populations to the cosmic reionization, and relevant physical mechanisms, will also be further investigated in separate papers.

\subsubsection{Chemical Evolution}
\label{sec:r3}

\begin{figure*}[t]
\begin{center}
\includegraphics[trim=0cm 0cm 0cm 0cm, clip, angle=0,width=1\textwidth]{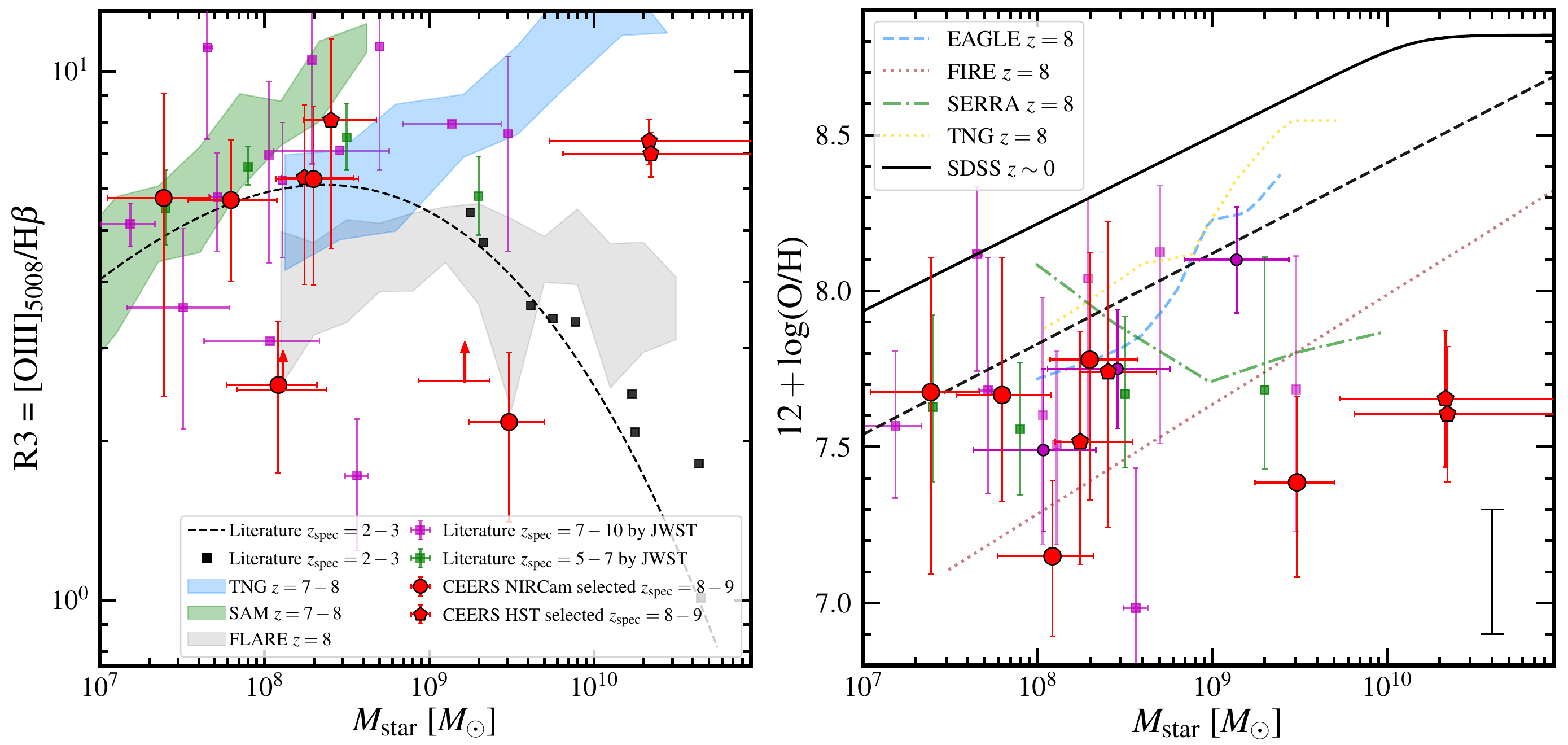}
\end{center}
\vspace{-0.4cm}
 \caption{
\textbf{Left:} 
R3 vs. $M_{\rm star}$ relation. 
The red circles and pentagons present our NIRCam- and \hst-selected galaxies at $z=8-9$ in CEERS, respectively, including lower limits.  
The magenta and green squares denote the recent \jwst\ results for lensed galaxies at $z=7.5$--9.5 \citep{trump2022, williams2022, heintz2022b,mascia2023} and general field galaxies at $z=5-7$ \citep{matthee2022}.  
The black squares show the results for general field galaxies at $z\sim2-3$ \citep{sanders2020,papovich2022}, where its R3--$M_{\rm star}$ relation is extrapolated by converting the mass--metallicity relation based on the R3 calibration of \cite{nakajima2022}. 
The color-shaded indicate the 16-84th percentile of simulated galaxies at $z=7-8$ in TNG50 and TNG100 \citep[e.g.,][]{nelson2019} and Santa Cruz SAM \citep[e.g.,][]{yung2019b, yung2022}, incorporating the nebular emission line calculations  \citep{hirschmann2022}, and FLARE \citep{wilkins2022a, wilkins2022b}. 
 \textbf{Right:}
Metallicity vs. $M_{\rm star}$ relation. 
Our and other R3 measurements at $z>7$ are shown in the left panel, converted using the R3-metallicity calibration, where the typical error scale from different calibrations is represented by the black bar in the lower right corner (see text).
The black lines are the best-fit relation at $z=0$ (solid) and $z=3$ (dashed) \citep{sanders2021}, while the other color lines denote the predictions from simulations \citep{schaye2015, ma2016, pallottini2022, ucci2023}.
The magenta circles are the measurements based on the \oiii4363 line for three lensed galaxies at $z=7.7$-8.5 \citep[e.g.,][]{trump2022, curti2022, schaerer2022}. 
The mass--metallicity relation at $z=8-9$ is likely placed below the relations at $z=0-3$, which is generally consistent with the predictions from simulations.  
\label{fig:r3-metallicity}}
\end{figure*}

The rest-frame optical emission lines have been the most intensively calibrated and widely exploited to constrain the metallicity in distant galaxies (see 
\citealt{maiolino2019} for a recent review).  
Our NIRSpec observations cover the wavelength of the key optical emission lines of \oiii5008, H$\beta$, and \oii3727. However, we do not securely detect any \oii\ lines from our $z_{\rm spec}$ sample. 
We thus focus on the R3 (\oiii/H$\beta$) measurement to examine the oxygen abundance in the following analysis, assuming a negligible contribution from \oii. 
The general agreement in the metallicity measurement between the R3-based method and the direct electron temperature method has been confirmed even at $z=5-8$ galaxies in recent \jwst\ studies based on the successful aural \oiii4363 line detection \citep[e.g.,][]{curti2022, matthee2022}.

In the left panel of Figure~\ref{fig:r3-metallicity}, we show our R3 measurements as a function of $M_{\rm star}$. We also measure and show the \hst-selected, luminous LBGs (Larson et al., in prep.) in the same manner\footnote{
One of the CEERS \hst-selected galaxies, MPT-ID1019, shows several possible AGN features \citep{larson2023}, and thus we do not include MPT-ID1019 in this analysis. 
}. 
These NIRCam- and \hst-selected samples enlarge the $M_{\rm star}$ parameter space and provide a unique opportunity to examine the mass--metallicity relation in a fair manner, owing to the homogeneous sample selection and follow-up sensitivity. 
For comparison, we also present the recent results for field galaxies at $z=2-3$ \citep{sanders2020,papovich2022} and $z=5-7$ \citep{matthee2022} and lensed galaxies at $z=7.5-9.5$ \citep{trump2022, williams2022, heintz2022b,mascia2023}. 
By implication, the R3--$M_{\rm star}$ relation at $z=2$ is extrapolated by converting the $z=2$ mass--metallicity relation \citep{sanders2020} based on the R3-metallicity calibration presented in \citet{nakajima2022}. 
We also show predictions from cosmological galaxy evolution models, incorporating the nebular emission line \citep{nelson2019, yung2019b, yung2022, hirschmann2022, wilkins2022a,wilkins2022b}. 

First, we find in our NIRCam- and \hst-selected samples that there is a potential increasing trend in R3 with increasing $M_{\rm star}$. 
This trend is different from the results at $z\sim2-3$ (black squares) but generally consistent with the model predictions (color-shaded regions), where the increasing trend is explained by the high ionization parameter in early galaxies (\citealt{hirschmann2022}; Hirschmann et al., in prep.). High O32 values of $\gtrsim10$ have been reported from the composite spectrum for $z\sim7.5$--8.0 galaxies \citep{sanders2023, tang2023, cameron2023a}, which also suggests the highly ionized ISM state in early galaxies. 
Second, we compare our measurements with recent \jwst\ results and find that our NIRCam-selected galaxies show the R3--$M_{\rm star}$ distribution similar to the lensed galaxies at $z=7.5-9.5$, ranging over the R3 values of $\sim2$--6 that are generally comparable or lower than those of the field galaxies at $z=5-7$. 
Below 12+$\log (O/H)\approx8$, the R3 values generally decrease towards low metallicity because the \oiii\ emission is suppressed in the low oxygen abundance \citep[e.g.,][]{maiolino2008, curti2017, bian2018, nakajima2022}. 
Thus, the lower R3 values in galaxies from CEERS and lensing samples at $z\gtrsim8$ with respect to the $z=5-7$ field galaxies might indicate a decreasing metallicity trend at a given $M_{\rm star}$ towards high redshifts. 
Although a few sources are placed much below the other sources and the model predictions at $\log(M_{\rm star}/M_{\odot})\gtrsim8.5$, this might indicate that the pristine gas inflow might be taking place in these exceptional galaxies, which maintains the oxygen abundance low and makes their \oiii\ line emissivity very low. 

In the right panel of Figure~\ref{fig:r3-metallicity}, 
we present the metallicity measurements at $z=0-9$.  
We estimate the metallicities of our sample using the R3 calibration of \cite{nakajima2022} that takes H$\beta$ EW ($\approx$ ionization state) into account and expands the calibration down to $12+\log(O/H) \simeq 6.9$ by analyzing the local extremely metal-poor galaxies. 
\footnote{
For the sources with EW(H$\beta$) $> 150{\rm \AA}$ and the other, we use the R3--metallicity conversion calibrated for high ($\gtrsim 200{\rm \AA}$) and low ($\lesssim 100{\rm \AA}$) H$\beta$ EW sources in \cite{nakajima2022}, respectively. 
In this conversion, we assume that our NIRCam-selected sample has $12+\log(O/H)\leq8.2$, because of their less massive properties ($\lesssim10^{9}M_{\odot}$, see Figure~\ref{fig:sfr-ms}) and lower limit of O32 (Table~\ref{tab:prop}) that correspond to a low oxygen abundance of $12+\log(O/H)\lesssim8.2$ based on the O32 calibration \citep[e.g.,][]{maiolino2008, curti2017, bian2018, perez-montero2021, nakajima2022, casey2022}.}
We obtain $12+\log(O/H)\simeq7.2-7.8$ for the NIRCam-selected galaxies at $z=8-9$. 
These estimates typically change by $\sim0.4$, if we use another calibration of \cite{bian2018}. 
Some \hst-selected galaxies exhibit high R3 values ($>6$) that surpass the possible range in the calibration mentioned above. In such cases, we employ an alternative calibration for simulated galaxies at $z=4-8$ (\citealt{hirschmann2022}; Hirschmann et al., in prep.)\footnote{
$R3=P0 + P1x + P2 x^{2}$,
where $P0 =-14.424, P1=3.521, P2 =-0.199$, and $x=12+\log(O/H)$} 
that reproduces the increasing R3 trend with increasing $M_{\rm star}$ in the left panel. 
This yields $12+\log(O/H)\simeq7.6-8.2$ for the \hst-selected galaxies. 
Based on these calibrations, we find that the galaxies at $z=8-9$ fall below the mass metallicity relations at $z=0-3$ \citep[e.g.,][]{sanders2021} extrapolated down to $M_{\rm star}\sim10^{7}\, M_{\odot}$. 
This is generally consistent with the model predictions \citep[e.g.,][]{schaye2015, ma2016, pallottini2022, ucci2023} and the recent \jwst\ results of the mass-metallicity measurements at $z=5-9.5$ \citep{matthee2022, heintz2022b, langeroodi2022}, but with different calibrations \citep[e.g.,][]{bian2018, curti2020}. 
By applying the R3 calibrations to these recent \jwst\ results in the same manner as above, we obtain $12+\log(O/H)\simeq7.7-8.1$ and $\simeq7.5-7.8$ for the $z=5-7$ field galaxies and the lensed galaxies at $z=6.9-8.1$, respectively. This implies that the metallicity of our NIRCam-selected galaxies at $z=8-9$ is consistent with the recent \jwst\ results within the large scatter. 
Combined with our SFR--$M_{\rm star}$ results (Section \ref{sec:sfr-mstar}), we find that galaxies at $\log(M_{\rm star}/M_{\odot})\simeq8-9$ have a trend of an increasing SFR and decreasing metallicity from $z\sim2-6$ to $z\sim8-9$.

\section{Summary}
\label{sec:summary}

In this paper, we present  \jwst\ NIRSpec multi-object spectroscopy results for 11 high-redshift galaxy candidates at $z\gtrsim8.5$ selected from deep NIRCam data taken in June 2022 in the Cosmic Evolution Early Release Science (CEERS) field. 
All targets are observed with the prism or medium-resolution G140M/F100LP, G235M/F170LP and G395M/F290LP gratings, both of which continuously covers $\sim$1--5~$\mu$m wavelengths in the observed frame, maximizing the chance of spectroscopic confirmation either by emission lines or the Lyman-$\alpha$ break. 
This is the first homogeneous, luminosity-selected follow-up \jwst\ spectroscopy for high-redshift galaxy candidates selected with NIRCam in the UV luminosity range of $M_{\rm UV}\in[-21,-18]$, setting the benchmark for future high-redshift galaxy selection at $z\gtrsim8.5$ with NIRCam. 
Comparing with the \hst-selected luminous ($M_{\rm UV}\simeq-22$) galaxies that are also spectroscopically confirmed at similar redshifts (Larson et al., in prep.), we also investigate the characteristics of these NIRCam-selected galaxies. The main findings of this paper are summarized as follows:

\begin{enumerate}
\item We systematically analyze 1D and 2D spectra and spectroscopically confirm 7 out of 11 targets with \oiii5008, 4960 lines at $z=7.762-8.998$. One of the candidates shows a potential single line detection at $5.20$~$\mu$m, which may indicate the galaxy redshift at $z=9.386$ or $z=9.697$, depending on whether the line is \oiii5008 or H$\beta$. We do not detect emission lines or Lyman-$\alpha$ continuum breaks in the other three candidates. 
\item Based on the original $z_{\rm phot}$ estimate, the success ratio of the spec-$z$ confirmation for $z_{\rm phot}=8.5-9.6$ candidates reaches $\simeq90$\% ($=6/7$). The absence of robust multiple-line identifications in three out of four $z_{\rm phot}>10$ candidates might indicate that strong optical emission lines (e.g., \oiii5008, H$\beta$) in these galaxies are redshifted beyond the NIRSpec wavelength range, and may support a high purity also in the $z_{\rm phot}>10$ candidate selection.  
The spec-$z$ confirmation results from the homogeneous sample selection and follow-up observations are widely useful to interpret the UV luminosity function studies now explored out to $z\sim17$ with \jwst/NIRCam. 
\item Although all $z_{\rm spec}$ values fall within the 95\% confidence interval of the original $z_{\rm phot}$ estimates, we find the majority of the sources show $z_{\rm spec}<z_{\rm phot}$, where almost half exceed the 68\% confidence interval. This may indicate that the galaxy templates generally used for high-redshift galaxy selection at $z\gtrsim8.5$ are still insufficient to recover their rest-frame UV color space. This could also be caused by the possible limitations inherent in using a single photo-$z$ code, the lack of a deep NIRCam dropout filter (e.g., F090W), or the potential bias due to the fact that the strong line identification is limited to $z\lesssim9.5$. 
\item We perform the SED fitting to the \hst+NIRCam photometry and \oiii5008 equivalent with (EW) and evaluate the star-formation rate (SFR) and stellar mass ($M_{\rm star}$) relation for the seven spec-$z$ confirmed sources. The typical SFR and $M_{\rm star}$ are estimated to be $\simeq4\,M_{\odot}$~yr$^{-1}$ and $\simeq$~10$^{8}\,M_{\odot}$. This yields a relatively high specific SFR of $\simeq40$~Gyr$^{-1}$ compared to the main sequence at $z=2-6$.  
\item We 
measure the rest-frame \oiii+H$\beta$ EW and obtain median values of 1100$^{+560}_{-730}{\rm \AA}$ and 570$^{+590}_{-410}{\rm \AA}$ for our NIRCam- and \hst-selected galaxies spectroscopically confirmed at $z=8-9$, respectively. This is much higher than typical star-forming galaxies at $z\sim1-2$ but consistent with previous photometric-based results at $z\sim7$ in a similar UV luminosity range. The potential larger EW(\oiii+H$\beta$) value in the NIRCam-selected galaxies may indicate these less luminous galaxies are not extremely metal-poor, but less metal-enriched systems with more ionizing photons maintaining the EW(\oiii+H$\beta$) slightly higher than the UV luminous \hst-selected galaxies. 
\item We also evaluate $\xi_{\rm ion}$ and find that our NIRCam- and \hst-selected galaxies have median values of $\log$(\xion/Hz~erg$^{-1}$) $=25.77^{+0.50}_{-0.43}$ and $25.47^{+0.33}_{-0.23}$, respectively, at $z_{\rm spec}=8-9$. These median estimates are much higher than typical star-forming galaxies at $z\sim1-2$ but consistent with the recent measurements for the \oiii\ emitters and Lyman-$\alpha$ emitters at $z=5-7$. The potential trend toward higher $\xi_{\rm ion}$ in less luminous galaxies indicates the important contributions from faint galaxies to the cosmic reionization, which is consistent with the empirical model predictions based on H$\alpha$ EW and UV luminosity.  
\item We analyze the \oiii/H$\beta$ line ratio ($=$ R3) as a function of $M_{\rm star}$ for the NIRCam- and \hst-selected galaxies. We find an increasing trend of R3 towards high $M_{\rm star}$. This trend is opposite to the $z\sim2-3$ results, while it is consistent with the predictions from the simulations because of the high ionization parameter in the early galaxies. 
With an empirical calibration of the R3 method, we evaluate the oxygen abundance and find that these $z=8-9$ galaxies have lower metallicity than $z=0-3$ galaxies at a given $M_{\rm star}$. This is generally consistent with the current galaxy formation and evolution models.   
\end{enumerate}

\begin{acknowledgments}
We thank Gabriel Brammer for sharing insights into the NIRSpec data analysis and Takashi Kojima and Ryan Endsley for useful discussions for this paper. We thank the entire CEERS team for their effort to design and execute this observational program, especially the work to design the MSA observations.  
This work is based on observations with the NASA/ESA/CSA James Webb Space Telescope obtained from the Mikulski Archive for Space Telescopes at the STScI, which is operated by the Association of Universities for Research in Astronomy (AURA), Incorporated, under NASA contract NAS5-03127.
We acknowledge support from NASA through STScI ERS award JWST-ERS-1345.
This research also made use of the NASA/IPAC Infrared Science Archive (IRSA), 
which is operated by the Jet Propulsion Laboratory, California Institute of Technology, under contract with the National Aeronautics and Space Administration. 
This project has received funding from NASA through the NASA Hubble Fellowship grant HST-HF2-51505.001-A awarded by the Space Telescope Science Institute, which is operated by the Association of Universities for Research in Astronomy, Incorporated, under NASA contract NAS5-26555.
\end{acknowledgments}

Some/all of the data presented in this paper were obtained from the Mikulski Archive for Space Telescopes (MAST) at the Space Telescope Science Institute. 
The specific observations analyzed can be accessed via \url{10.17909/z7p0-8481}.

\software{ \texttt{Astropy} \citep{astropy2013},  
\texttt{CIGALE} \citep{boquien2019}, 
\texttt{Mosviz} \citep{developers2023}. 
}

\bibliographystyle{apj}
\bibliography{apj-jour,reference}

\appendix

\section{Sources without spectroscopic redshift confirmation}

\begin{figure*}[t]
\includegraphics[trim=0.2cm 0cm 0cm 0cm, clip, angle=0,width=1\textwidth]{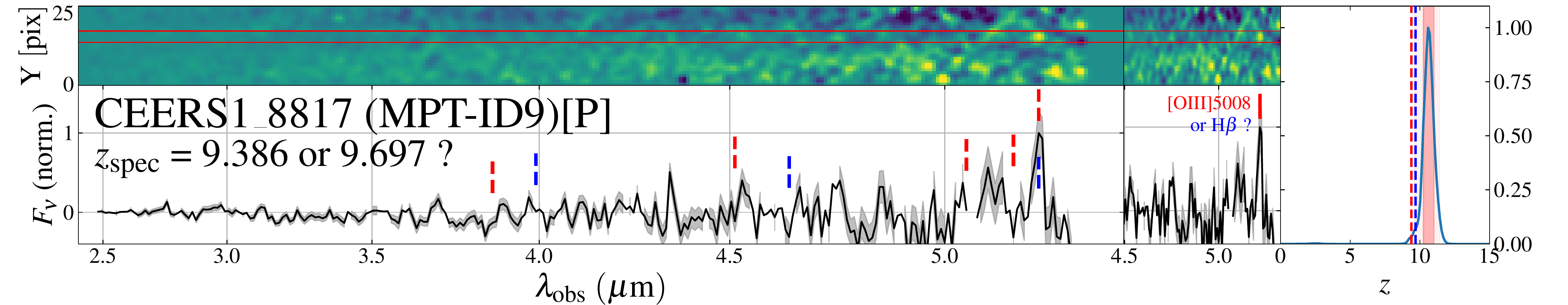}
 \caption{
 Same as Figure~\ref{fig:spectra}, but for MPT-ID9 with the potential single line detection at $5.20$~$\mu$m. The red and blue vertical lines indicate the expected wavelengths for the \oiii5008, \oiii4969, H$\beta$, H$\gamma$, and \oii3727, from right to left when the 5.20~$\mu$m line feature corresponds to \oiii5008 and H$\beta$, respectively. For the potential line, we rule out the possibility of \oiii4960, because the 3$\times$ more bright \oiii5008 line should be detected at the edge of the spectrum in this case. 
The \oii3727 at $z=12.95$ is also unlikely based on the photometric redshift $P(z)$. 
\label{fig:ID9}}
\end{figure*}

\begin{figure*}
\includegraphics[trim=0.2cm 0cm 0cm 0cm, clip, angle=0,width=1\textwidth]{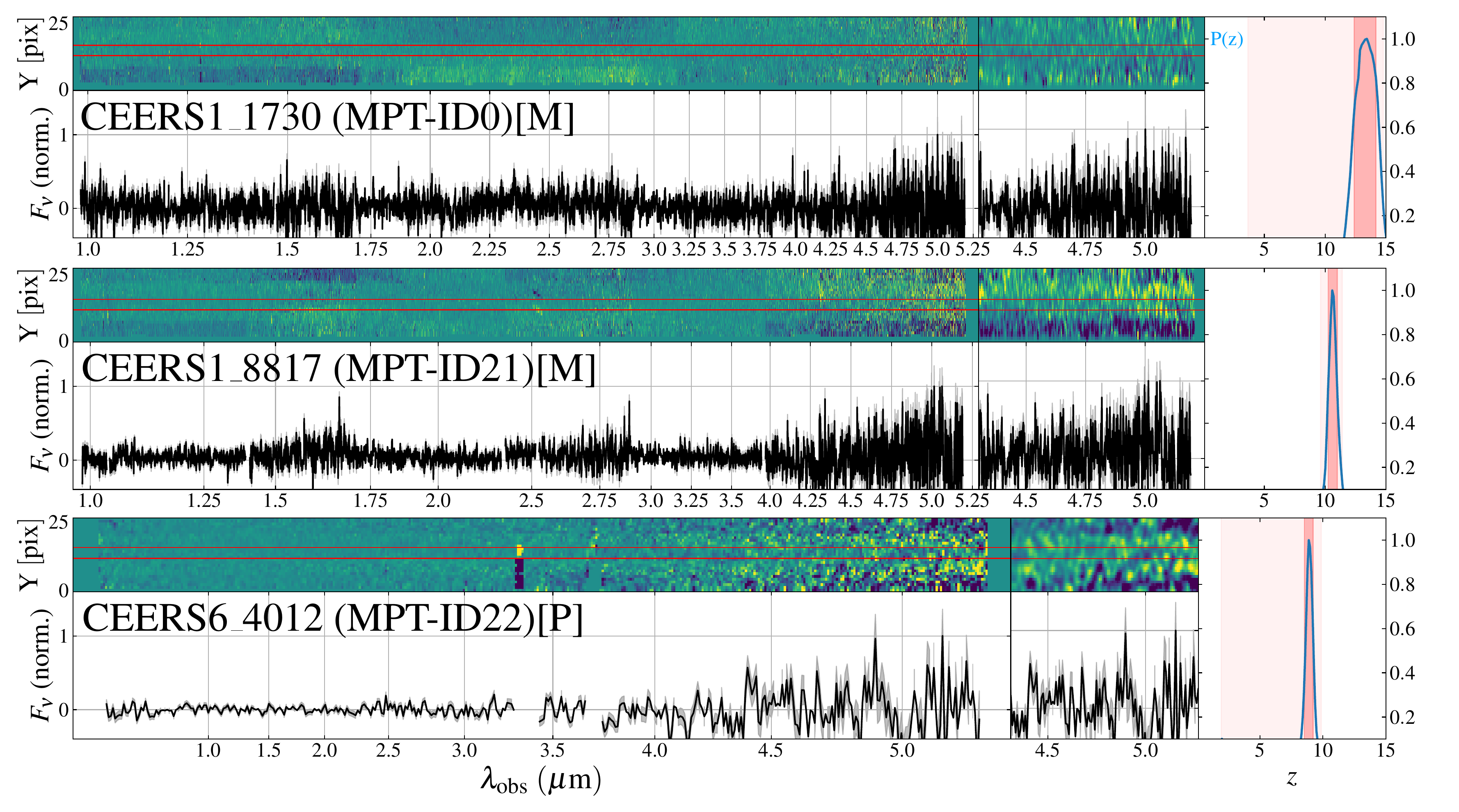}
 \caption{
Same as Figure~\ref{fig:spectra}, but for MPT-ID0, 21, and 22 without a robust line or continuum detection. 
\label{fig:non-detect}}
\end{figure*}

In Figure \ref{fig:ID9} and \ref{fig:non-detect}, we show the 1D+2D spectra for the sources that are not spectroscopically confirmed in this study. 

\section{Continuum break}

\begin{figure}
\includegraphics[trim=0cm 0cm 0cm 0cm, clip, angle=0,width=0.5\textwidth]{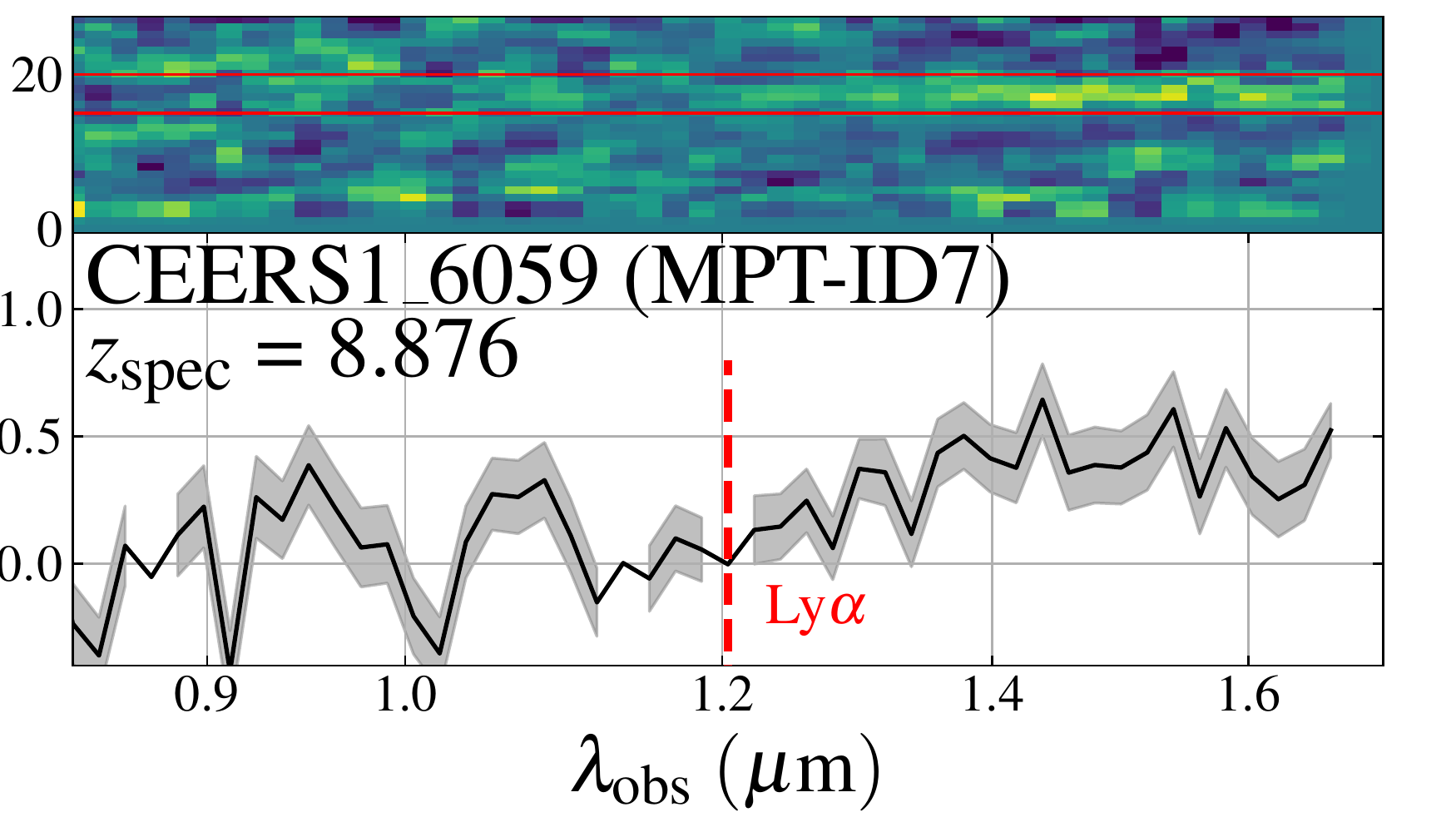}
 \caption{
 Same as Figure~\ref{fig:spectra}, but zoomed at around $\sim1\mu$m of MPT-ID7 whose Lyman-$\alpha$ continuum break is identified in the 1D and 2D spectra. The wavelength of the continuum break is consistent with the emission-line-determined spectroscopic redshift.  
\label{fig:id7_break}}
\end{figure}

In Figure \ref{fig:id7_break}, we show the 1D+2D prism spectrum for MPT-ID7 
in the spectral region around $1\mu$m.
We detect continuum emission at $\lambda \gtrsim 1.2\mu$m consistent with expectations for the Ly$\alpha$ break at the redshift determined from the \oiii and H$\beta$ emission lines.

\section{Best-fit SED}

\begin{figure}
\begin{center}
\includegraphics[trim=0cm 0cm 0cm 0cm, clip, angle=0,width=0.48\textwidth]{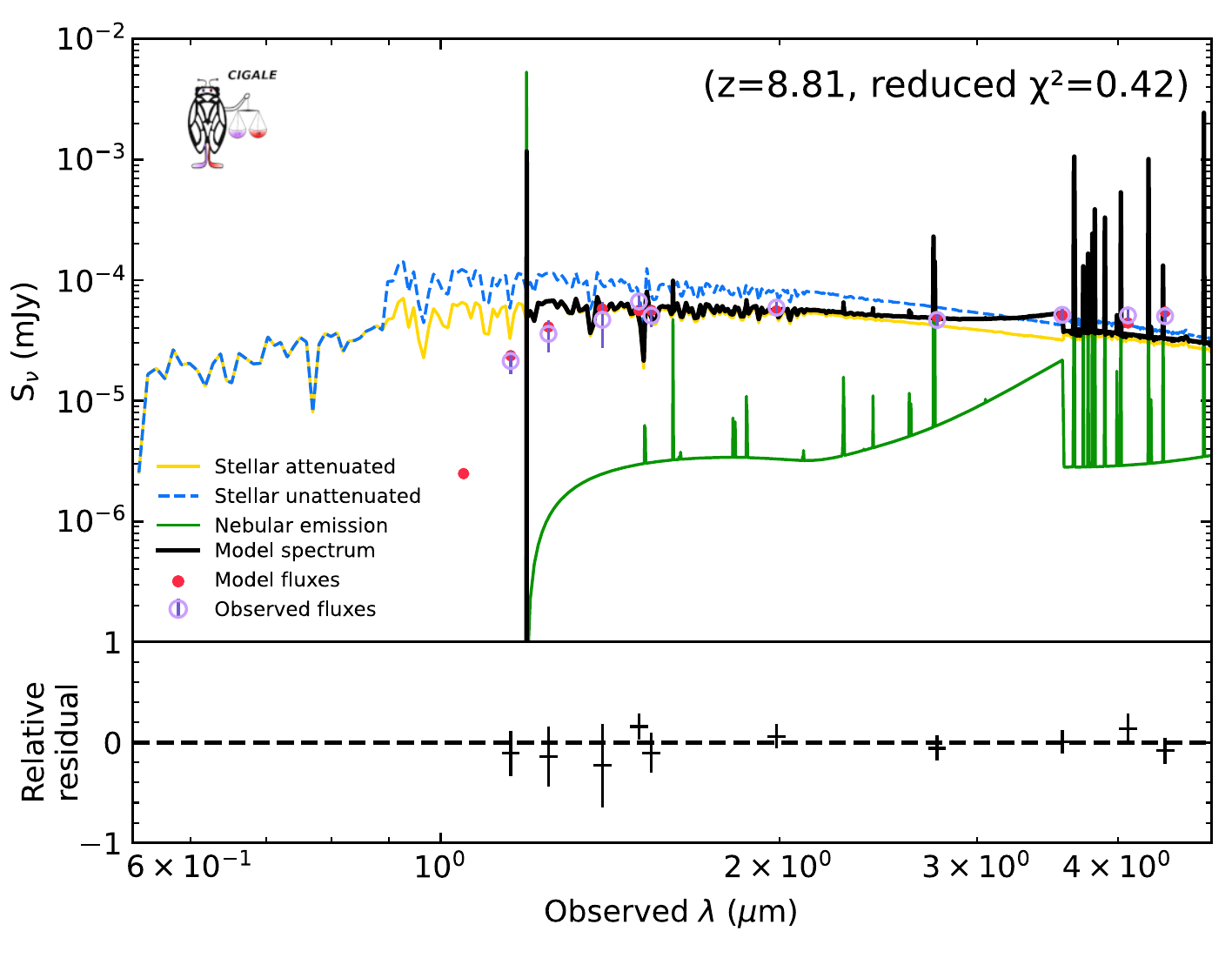}
\end{center}
 \caption{
Best-fit SED for MPT-ID2 modeled with {\sc cigale}.
Colored curves represent the emission from the various components described in the label, with the black curve representing their sum.
Magenta and red circles indicate the observed and model-predicted fluxes in the NIRCam broadband filters, respectively. }
\label{fig:sed}
\end{figure}

In Figure \ref{fig:sed}, we show our best-fit SED model for MPT-ID2 by using {\sc cigale} as an example. A comprehensive list and further details of the SED models will be provided in Burgarella et al. (in preparation).

\end{document}